\documentclass{amsart}%
\usepackage{amsfonts}
\usepackage{amsmath}
\usepackage{amssymb}
\usepackage{graphicx}%
\theoremstyle{plain}
\newtheorem{acknowledgement}{Acknowledgement}

\newtheorem{definition}{Definition}

\newtheorem{notation}{Notation}

\numberwithin{equation}{section}
\begin{document}
\title[Non-Archimedian Replicator Dynamics]{Non-Archimedean Replicator Dynamics and Eigen's Paradox}
\author{W. A. Z\'{u}\~{n}iga-Galindo}
\address{Centro de Investigaci\'{o}n y de Estudios Avanzados del Instituto
Polit\'{e}cnico Nacional\\
Departamento de Matem\'{a}ticas, Unidad Quer\'{e}taro\\
Libramiento Norponiente \#2000, Fracc. Real de Juriquilla. Santiago de
Quer\'{e}taro, Qro. 76230\\
M\'{e}xico. }
\email{wazuniga@math.cinvestav.edu.mx}
\thanks{The author was partially supported by Conacyt Grant No. 250845.}
\date{25/01/2018}

\begin{abstract}
We present a new non-Archimedean model of evolutionary dyna\-mics, in which
the genomes are represented by $p$-adic numbers. In this model the genomes
have a variable length, not necessarily bounded, in contrast with the
classical models where the length is fixed. The time evolution of the
concentration of a given genome is controlled by a $p$-adic evolution
equation. This equation depends on a fitness function $f$ and on mutation
measure $Q$. By choosing a mutation measure of Gibbs type, and by using a
$p$-adic version of the Maynard Smith Ansatz, we show the existence of
threshold function $M_{c}(f,Q)$, such that the long term survival of a genome
requires that its length grows faster than $M_{c}(f,Q)$. This implies that
Eigen's paradox does not occur if the complexity of genomes grows at the right
pace. About twenty years ago, Scheuring and Poole, Jeffares, Penny proposed a
hypothesis to explain Eigen's paradox. Our mathematical model shows that this
biological hypothesis is feasible, but it requires $p$-adic analysis instead
of real analysis. More exactly, the Darwin-Eigen cycle proposed by Poole et
al. takes place if the length of the genomes exceeds $M_{c}(f,Q)$.

\end{abstract}
\maketitle

\section{Introduction}

In this article we present a new non-Archimedean model of evolutionary
dyna\-mics of replicating single-stranded RNA genomes, which constitutes a
non-Archi\-medean generalization of the classical Eigen quasispecies model.
Mathematically speaking, the new model is a class of $p$-adic
pseudodifferential evolution equations, which depends on a fitness function
$f$ and on a mutation measure $Qdx$. In the new model a sequence (genome) is
specified by a $p$-adic number:%
\begin{equation}
x=x_{-k}p^{-k}+x_{-k+1}p^{-k+1}+\ldots+x_{0}+x_{1}p+\ldots,\text{ with }%
x_{-k}\neq0\text{,} \label{p-adic-number}%
\end{equation}
where $p$ denotes a fixed prime number, and the $x_{j}$s \ are $p$-adic
digits, i.e. numbers in the set $\left\{  0,1,\ldots,p-1\right\}  $. Thus, in
our model the sequences may have arbitrary length. In the case $p=2$, we
obtain the set of binary sequences of arbitrary length. The set of all
possible sequences constitutes the field of $p$-adic numbers $\mathbb{Q}_{p}$.
There are natural field operations, sum and multiplication, on series of form
(\ref{p-adic-number}), see e.g. \cite{Koblitz}. There is also a natural norm
in $\mathbb{Q}_{p}$ defined as $\left\vert x\right\vert _{p}=p^{k}$, for a
nonzero $p$-adic number $x$ of the form (\ref{p-adic-number}). The field of
$p$-adic numbers with the distance induced by $\left\vert \cdot\right\vert
_{p}$ is a complete ultrametric space. The ultrametric property refers to the
fact that $\left\vert x-y\right\vert _{p}\leq\max\left\{  \left\vert
x-z\right\vert _{p},\left\vert z-y\right\vert _{p}\right\}  $ for any $x$,
$y$, $z$ in $\mathbb{Q}_{p}$.

We denote by $\mathbb{Z}_{p}$ the unit ball, which consists of the all the
sequences with expansions of the form (\ref{p-adic-number}) with $k\geq0$, and
by $\mathbb{Z}_{p}^{\times}$, the subset of $\mathbb{Z}_{p}$ consisting of the
$p$-adic numbers with norm $1$. This last set is the disjoint union of sets of
the form $j+p\mathbb{Z}_{p}$, for $j\in\left\{  1,\ldots,p-1\right\}  $. Each
set of the form $j+p\mathbb{Z}_{p}$ is (in a natural form) an infinite rooted
tree. Then, all the sequences contained in the set $j+p\mathbb{Z}_{p}$ are
naturally organized in a `phylogenetic tree', and the set $\mathbb{Z}%
_{p}^{\times}$ is a forest formed by the disjoint union of $p-1$ infinite
rooted trees. On the other hand, $\mathbb{Q}_{p}\smallsetminus\left\{
0\right\}  $ is a countable disjoint union of scaled versions of the forest
$\mathbb{Z}_{p}^{\times}$, more precisely, $\mathbb{Q}_{p}\smallsetminus
\left\{  0\right\}  =%
{\textstyle\bigsqcup\nolimits_{k=-\infty}^{k=+\infty}}
p^{k}\mathbb{Z}_{p}^{\times}$. The field of $p$-adic numbers has a fractal
structure, see e.g. \cite{Alberio et al}, \cite{V-V-Z}. For `pictures' of
\ $p$-adic spaces the reader may consult \cite{Holly}.

Our model of evolutionary dynamics is a $p$-adic continuous model. The fitness
landscape is given by a function $f:$ $\mathbb{Q}_{p}\rightarrow\mathbb{R}%
_{+}$. In this article, we assume that $f$ is a test function, which means
that $f$ is a locally constant function with compact support. With respect to
the mutation mechanism, we only assume the existence of a mutation measure
$Q\left(  \left\vert x\right\vert _{p}\right)  dx$, where $Q:$ $\mathbb{R}%
_{+}\rightarrow\mathbb{R}_{+}$, $dx$ is the normalized Haar measure of the
group $\left(  \mathbb{Q}_{p},+\right)  $, and $\int Q\left(  \left\vert
x\right\vert _{p}\right)  dx=1$, such that the probability that a sequence $x$
mutates into a sequence \ belonging to the set $B$ is given by $\int_{B}$
$Q\left(  \left\vert x-y\right\vert _{p}\right)  dy$. \ In our model the
concentration $X\left(  x,t\right)  $ of the sequence $x$ at the time $t$ is
controlled by the following evolution equation:%
\begin{equation}
\frac{\partial X\left(  x,t\right)  }{\partial t}=Q\left(  \left\vert
x\right\vert _{p}\right)  \ast\left\{  f\left(  \left\vert x\right\vert
_{p}\right)  X\left(  x,t\right)  \right\}  -\Phi\left(  t\right)  X\left(
x,t\right)  , \label{EC_0}%
\end{equation}
where $\Phi\left(  t\right)  =\int_{\mathbb{Q}_{p}}f\left(  \left\vert
y\right\vert _{p}\right)  X\left(  y,t\right)  dy$.

The term $Q\left(  \left\vert x\right\vert _{p}\right)  \ast\left\{  f\left(
\left\vert x\right\vert _{p}\right)  X\left(  x,t\right)  \right\}  $
\ represents the rate at which the sequences are mutating into the sequence
$x$. We now assume that the replication reactions occur in a chemostat, see
\cite{Tannembaum et al} and the references therein, which is a device that
allows the maintenance of a constant population size, this corresponds to the
term $-\Phi\left(  t\right)  X\left(  x,t\right)  $. From this discussion, we
conclude that the instantaneous change of the concentration $X\left(
x,t\right)  $ is given by (\ref{EC_0}).

The mathematical study of the equation (\ref{EC_0}), and in particular of the
associated Cauchy problem, is an open problem. The functions $f$ and $Q$ may
depend on time, but here we study only the Cauchy problem associated to
(\ref{EC_0}) in a particular case, by using $p$-adic wavelets, see Section
\ref{Section_Cauchy_problem}.

A central problem in the origin of life is the reproduction of primitive
organisms with sufficient fidelity to maintain the information coded in the
primitive genomes. In the case of sequences (genomes) with constant
length,\ and under the assumption of independent point mutations, that is,
assuming that during the replication process each digit (nucleotide) has a
fixed probability of being replaced for another digit, and that this
probability is independent \ of all other digits, Eigen discovered that the
mutation process places a limit on the number of digits that a genome may
have, see e.g. \cite{Eigen1971}, \cite{Eigen et al}, \cite{Nowak},
\cite{SchusterPeter}, \cite{Tannembaum et al}. This critical size is called
the error threshold of replication. The genomes larger than this error
threshold will be unable to copy themselves with sufficiently fidelity, and
the mutation process will destroy the information in subsequent generations of
these genomes. This discussion naturally drives to the following question: how
is it possible the existence of large stable living organisms on earth? To
create more complex organisms (that is to have more genetic complexity), it is
necessary to encode more information in larger genomes by using a replication
mechanism with greater fidelity. But the information for creating
error-correcting mechanisms (enzymes) should be encoded in the genomes, which
have a limited size. Hence, we arrive to the `Cath-22' or Eigen's paradox of
the origin of life: \textquotedblleft no large genome without enzymes, and no
enzymes without a large genome,\textquotedblright\ see \cite[p. 317]{Maynad
Smith}, \cite{Szat-tree}. Then main consequence of our $p$-adic model of
evolution is that it gives a completely\ `new mathematical' perspective of the
Eigen paradox.

We show that there is a finite ultrametric space $G_{M}$ consisting of
$p$-adic sequences of finite length $2M$, which is a rooted tree with $2M+1$
levels and $p^{2M}$ branches at the top level, and with a distance induced by
the restriction of the $p$-adic norm to $G_{M}$, such that the equation
(\ref{EC_0})\ admits a discretization\ (a finite approximation) of the form
\begin{equation}
\frac{d}{dt}X\left(  J,t\right)  =\frac{1}{C}\sum_{I\in G_{M}}Q\left(
\left\vert J-I\right\vert _{p}\right)  f\left(  \left\vert I\right\vert
_{p}\right)  X(I,t)-\Phi_{M}\left(  t\right)  X\left(  J,t\right)  \text{
\ for }J\in G_{M}, \label{Eigen-discrete1}%
\end{equation}
which is exactly the Eigen model on $G_{M}$, see Section \ref{Sect_DIs}. In
the classical Eigen model the space of sequences is a finite metric space with
a distance induced by the Hamming weight, in our model this space is replaced
by $G_{M}$. The main restriction of this last model is that the mutation
matrix is a function of the $p$-adic distance between two sequences.

Heuristically speaking, the limit when $M$ tends to infinity of the Eigen
system (\ref{Eigen-discrete1}) is the evolution equation (\ref{EC_0}). This
heuristics can be justified by using the techniques from
\cite{zuniga-Nonlieal} and the references therein, more exactly, we have that
the solutions of the Cauchy problem associated to (\ref{EC_0}) can be very
well approximated by solutions of the Cauchy problem associated to
(\ref{Eigen-discrete1}), in a suitable function space, when $M$ tends to infinity.

In \cite{Maynad Smith}, Maynard Smith introduced a mathematical approximation
(the Maynard Smith ansatz) that allows to study the error threshold problem
without solving the original Eigen system. This ansatz can be extended to the
$p$-adic setting, see Section \ref{Section-Maynard-Smith}. In this
approximation the space of sequences is divided into two disjoint groups, each
of them with a fix fitness, say $a$ and $b$, with $a>b$. The ansatz provides
an inequality which gives a necessary and sufficient condition for the long
term survival of the group of sequences with fitness $a$. It is remarkable
that this inequality is the classical one, see Section
\ref{Section-Maynard-Smith}. We use this ansatz to study the Eigen paradox for
two different families of mutation measures. The measures of the first family
are supported in the unit ball, the second is a family of Gibbs measures. In
both cases we show that the Eigen paradox does not occur if the length of the
sequences grows to the right pace. We propose using a Gibbs type measure:
$\frac{1}{C\left(  \alpha,\beta\right)  }e^{-\beta\left\vert x\right\vert
_{p}^{\alpha}}dx$, where $\alpha$, $\beta$ are positive constants,
and\ $C\left(  \alpha,\beta\right)  $\ is a normalization constant. The two
main reasons for this choice are: first, $\left\vert x\right\vert _{p}%
^{\alpha}$ is the simplest energy function which depends on $\log
_{p}\left\vert x\right\vert _{p}$, the `$p$-adic length' of the sequence, and
second, the discretization of the $p$-adic replicator equation (\ref{EC_0})
attached to this mutation measure is connected with the matrices defining
certain Markov chains, that typically are used in models of molecular
evolution, see e.g. \cite{L-N}.

The $p$-adic version of the Maynard Smith ansatz provides a necessary and
sufficient condition for the long term survival of a group of sequences of the
form $I+p^{M}\mathbb{Z}_{p}$, where $I$ is an infinite sequence which plays
the role of the master sequence, which are in competition with the group of
sequences $\mathbb{Q}_{p}\smallsetminus\left[  I+p^{M}\mathbb{Z}_{p}\right]
$. We establish the existence of an \textit{error threshold function }%
$M_{c}(s,\alpha,\beta)$, which depends on $\ln s$, with $1-s=\frac{b}{a}$, and
with $f\mid_{I+p^{M}\mathbb{Z}_{p}}\equiv a>f\mid_{\mathbb{Q}_{p}%
\smallsetminus I+p^{M}\mathbb{Z}_{p}}\equiv b$, such that the long term
survival of the sequences in the group $I+p^{M}\mathbb{Z}_{p}$ requires that
$M>M_{c}(s,\alpha,\beta)$. This means that under a `fierce competition'
between the two groups (i.e. when $s\rightarrow0^{+}$), the long term survival
of the first group requires that all the sequences in this group approaches to
$I$, in such way that the logarithm of the $p$-adic norm of the difference
\ of any of these sequences and $I$ is greater than $M_{c}(s,\alpha,\beta)$.
Notice that the set of sequences of finite length in $I+p^{M}\mathbb{Z}_{p}$
is a dense subset, the mentioned condition implies that the long term survival
of these sequences requires that the length of them grow.

On the other hand, if $M$ is upper bounded, then $M\leq M_{c}(s,\alpha,\beta)$
for $s$ sufficiently small, which is a version of the classical threshold
condition. In conclusion, our $p$-adic model of evolution predicts that
Eigen's paradox does not occur if the complexity of the genomes grow at the
right pace. About twenty years ago, Scheuring \cite{Sch} and Poole et al.
\cite{Poole etal} proposed a hypothesis to explain Eigen's paradox. Our
mathematical model gives life to this biological hypothesis, more precisely,
\textit{the Darwin-Eigen cycle}\ proposed in \cite{Poole etal} takes place
under the condition \ $M>M_{c}(s,\alpha,\beta)$: larger genome size improves
the replication fidelity, and this in turn increases the Eigen limit on the
length of the genome, which allows the evolution of larger genome size. In
turn, this allows the evolution of new function, which could further improve
the replication fidelity, and so on. See Section
\ref{Section_error_T_discussion}\ for an in-depth discussion about this matter.

In Section \ref{Section_Cauchy_problem}, we study the Cauchy problem attached
to the $p$-adic replicator equation, in the case in which the initial
concentration and the fitness are test functions. By using $p$-adic wavelets
and the classical method of separation of variables, we show the existence of
a solution $X(x,t)$ for the mentioned Cauchy problem. Then we show that
$X(x)=\lim_{t\rightarrow+\infty}$\ $X(x,t)$ exists, and it is a probability
density concentrated in the support of the fitness function. This steady state
concentration is the $p$-adic counterpart of the classical quasispecies. It is
controlled by fitness function and by the largest eigenvalue of the operator%
\[
\boldsymbol{W}\varphi\left(  x\right)  =Q(\left\vert x\right\vert _{p}%
)\ast\left\{  f\left(  \left\vert x\right\vert _{p}\right)  \varphi\left(
x\right)  \right\}
\]
for $\varphi$\ supported in a finite union of disjoint balls. The $p$-adic
quasispecies behave \ entirely different \ to the classical ones. \ An
in-depth understanding \ of the $p$-adic quasispecies require developing of
numerical methods for $p$-adic evolution equations.

An ultrametric space $(M,d)$ is a metric space $M$ with a distance satisfying
$d(A,B)\leq\max\left\{  d\left(  A,C\right)  ,d\left(  B,C\right)  \right\}  $
for any three points $A$, $B$, $C$ in $M$. In the middle of the 80s the idea
of using ultrametric spaces to describe the states of complex biological
systems, which naturally possess a hierarchical structure, emerged in the
works of Frauenfelder, Parisi, Stein, among others, see e.g. \cite{Dra-Kh-K-V}%
, \cite{Fraunfelder et al}, \cite{M-P-V}, \cite{R-T-V}. Frauenfelder et al.
proposed, based on experimental data, that the space of states of certain
proteins have an ultrametric structure, \cite{Fraunfelder et al}. Mezard,
Parisi, Sourlas and Virasoro discovered, in the context of the mean-field
theory of spin glasses, that the space of states of such systems has an
ultrametric structure, see e.g. \cite{M-P-V}, \cite{R-T-V}. A central paradigm
in physics of complex systems (for instance proteins) asserts that the
dynamics of such systems can be modeled as a random walk in the energy
landscape of the system, see e.g. \cite{Fraunfelder et al}, \cite{KKZuniga},
\cite{Kozyrev SV}, and the references therein. In this framework, the energy
landscape \ of a complex system is approximated by a pair consisting of an
ultrametric space and a function on this space describing the distribution of
the activation barriers, see e.g. \cite{Becker et al}. The dynamics of a such
system can is described by a system of equations of type%
\begin{equation}
\frac{\partial u\left(  i,t\right)  }{\partial t}=\sum_{j\neq i}J\left(
j,i\right)  v(j)u\left(  j,t\right)  -\sum_{j\neq i}J\left(  i,j\right)
v\left(  i\right)  u\left(  i,t\right)  ,\text{ }i=1,\ldots,N,
\label{Eq_System_0}%
\end{equation}
where the indices $i$, $j$\ number the states of the system (which correspond
to local minima of energy), $u(i,t)$ denotes the concentration of particles at
the state $i$ and at time $t$, $J\left(  i,j\right)  \geq0$\ is the
probability per unit time (or transition rate) of a transition from $i$\ to
$j$, and the $v(j)>0$\ are the basin volumes.\ We now assume that the space of
states of the system is a finite ultrametric space $(M,\left\vert
\cdot\right\vert )$, where the distance comes from a norm $\left\vert
\cdot\right\vert $, and that $J\left(  j,i\right)  =J\left(  \left\vert
j-i\right\vert \right)  $, $v(j)=v(\left\vert j\right\vert )$, $\sum
_{j}J\left(  \left\vert j-i\right\vert \right)  =1$, then the master equations
(\ref{Eq_System_0}) take the form%
\begin{equation}
\frac{\partial u\left(  i,t\right)  }{\partial t}=\sum_{j\neq i}J\left(
\left\vert j-i\right\vert \right)  v(\left\vert j\right\vert )u\left(
j,t\right)  -\left(  1-J(0)\right)  v(\left\vert i\right\vert )\left(
i,t\right)  \text{ \ for }i\in X\text{.} \label{Eq_1_A}%
\end{equation}
This system of equations is `similar' but not equal to the ultrametric version
of Eigen's system of equations up to the term $-\Phi\left(  t\right)  u\left(
i,t\right)  $, see (\ref{Eigen-discrete1}). The difference is that in
(\ref{Eq_1_A}) there are no transitions from state $i$ into $i$, and such
transitions play an important role in the Eigen ultrametric system
(\ref{Eigen-discrete1}).

It is relevant to mention that the kinetic models (\ref{Eq_System_0}) are
deeply connected with the theory of spin glasses, see e.g. \cite{Av-8},
\cite{KKZuniga}, \cite{Parisi-Sourlas}, \cite{R-T-V}, and the references
therein, \ and that techniques from this area have been successfully used in
the study of Eigen's model, see e.g. \cite{Let}, \cite{Park-Deem},
\cite{Saakian1}-\cite{Saakian3}.

In \cite{Av-Zhu}, see also \cite{Av-Zhu-2}, Avetisov and Zhuravlev proposed
using the one-dimensional ultrametric diffusion equation in the theory of
biological evolution. This equation (also known as the $p$-adic heat equation)
was introduced by Vladimirov, Volovich and Zelenov in \cite{V-V-Z}. A very
general theory of such equations is now available see e.g. \cite{Koch},
\cite{KKZuniga}, \cite{Zuniga-LNM-2016}. In \cite{Av-Zhu}, the authors
proposed that an evolutionary model based on the ultrametric diffusion
equation corresponds to the hierarchical picture of the referent description
of the biological world, and that the problem of error catastrophe has a
natural solution in the proposed framework. This approach does not allow to
analyze directly the error catastrophe in the usual sense.

In the last thirty years there has been a strong interest in the developing of
ultrametric models in biology, see for instance, \ \cite{Asano-Khrennikov2}%
-\cite{Dubichar-Khrennikov}, \cite{Fraunfelder et al}, \cite{Khrennikov1}%
-\cite{KKZuniga}, \cite{Kozyrev SV}-\cite{Kozyrev-Khrennikov}. The results
presented in this article are framed in this development and they confirm the
relevance of the ultrametricity in modeling biological systems which have
natural hierarchical structures.

We did our best to write a self-contained article addressed to a general
audience (biologists, mathematicians, physicists, among others). Numerical
simulations of our model are a not a straightforward matter due to several new
features, among them, the problem of visualization of $p$-adic objects.

\section{\label{Fourier Analysis}$p$-Adic Analysis: Essential Ideas}

\subsection{The field of $p$-adic numbers}

Along this article $p$ will denote a prime number. The field of $p-$adic
numbers $%
\mathbb{Q}
_{p}$ is defined as the completion of the field of rational numbers
$\mathbb{Q}$ with respect to the $p-$adic norm $|\cdot|_{p}$, which is defined
as
\[
\left\vert x\right\vert _{p}=\left\{
\begin{array}
[c]{lll}%
0 & \text{if} & x=0\\
&  & \\
p^{-\gamma} & \text{if} & x=p^{\gamma}\frac{a}{b}\text{,}%
\end{array}
\right.
\]
where $a$ and $b$ are integers coprime with $p$. The integer $\gamma:=ord(x)
$, with $ord(0):=+\infty$, is called the\textit{\ }$p-$\textit{adic order of}
$x$.

Any $p-$adic number $x\neq0$ has a unique expansion of the form
\[
x=p^{ord(x)}\sum_{j=0}^{\infty}x_{j}p^{j},
\]
where $x_{j}\in\{0,\dots,p-1\}$ and $x_{0}\neq0$. By using this expansion, we
define \textit{the fractional part of }$x\in\mathbb{Q}_{p}$, denoted
$\{x\}_{p}$, as the rational number
\[
\left\{  x\right\}  _{p}=\left\{
\begin{array}
[c]{lll}%
0 & \text{if} & x=0\text{ or }ord(x)\geq0\\
&  & \\
p^{ord(x)}\sum_{j=0}^{-ord_{p}(x)-1}x_{j}p^{j} & \text{if} & ord(x)<0.
\end{array}
\right.
\]
In addition, any non-zero $p-$adic number can be represented uniquely as
$x=p^{ord(x)}ac\left(  x\right)  $ where $ac\left(  x\right)  =\sum
_{j=0}^{\infty}x_{j}p^{j}$, $x_{0}\neq0$, is called the \textit{angular
component} of $x$. Notice that $\left\vert ac\left(  x\right)  \right\vert
_{p}=1$.

For $r\in\mathbb{Z}$, denote by $B_{r}(a)=\{x\in%
\mathbb{Q}
_{p};\left\vert x-a\right\vert _{p}\leq p^{r}\}$ \textit{the ball of radius
}$p^{r}$ \textit{with center at} $a\in%
\mathbb{Q}
_{p}$, and take $B_{r}(0):=B_{r}$. The ball $B_{0}$ equals $\mathbb{Z}_{p}$,
\textit{the ring of }$p-$\textit{adic integers of }$%
\mathbb{Q}
_{p}$. We also denote by $S_{r}(a)=\{x\in\mathbb{Q}_{p};|x-a|_{p}=p^{r}\}$
\textit{the sphere of radius }$p^{r}$ \textit{with center at} $a\in%
\mathbb{Q}
_{p}$, and take $S_{r}(0):=S_{r}$. We notice that $S_{0}^{1}=\mathbb{Z}%
_{p}^{\times}$ (the group of units of $\mathbb{Z}_{p}$). The balls and spheres
are both open and closed subsets in $%
\mathbb{Q}
_{p}$. In addition, two balls in $%
\mathbb{Q}
_{p}$ are either disjoint or one is contained in the other.

The metric space $\left(
\mathbb{Q}
_{p},\left\vert \cdot\right\vert _{p}\right)  $ is a complete ultrametric
space. As a topological space $\left(
\mathbb{Q}
_{p},|\cdot|_{p}\right)  $ is totally disconnected, i.e. the only connected
\ subsets of $%
\mathbb{Q}
_{p}$ are the empty set and the points. In addition, $\mathbb{Q}_{p}$\ is
homeomorphic to a Cantor-like subset of the real line, see e.g. \cite{Alberio
et al}, \cite{V-V-Z}. A subset of $\mathbb{Q}_{p}$ is compact if and only if
it is closed and bounded in $\mathbb{Q}_{p}$, see e.g. \cite[Section
1.3]{V-V-Z}, or \cite[Section 1.8]{Alberio et al}. The balls and spheres are
compact subsets. Thus $\left(
\mathbb{Q}
_{p},|\cdot|_{p}\right)  $ is a locally compact topological space.

\begin{notation}
We will use $\Omega\left(  p^{-r}|x-a|_{p}\right)  $ to denote the
characteristic function of the ball $B_{r}(a)$. We will use the notation
$1_{A}$ for the characteristic function of a set $A$.
\end{notation}

\subsection{Some function spaces}

A complex-valued function $\varphi$ defined on $%
\mathbb{Q}
_{p}$ is \textit{called locally constant} if for any $x\in%
\mathbb{Q}
_{p}$ there exist an integer $l(x)\in\mathbb{Z}$ such that
\begin{equation}
\varphi(x+x^{\prime})=\varphi(x)\text{ for }x^{\prime}\in B_{l(x)}.
\label{local_constancy_parameter}%
\end{equation}
\ A function $\varphi:%
\mathbb{Q}
_{p}\rightarrow\mathbb{C}$ is called a \textit{Bruhat-Schwartz function (or a
test function)} if it is locally constant with compact support. In this case,
we can take $l=l(\varphi)$ in (\ref{local_constancy_parameter}) independent of
$x$, the largest of such integers is called \textit{the parameter of local
constancy} of $\varphi$. The $\mathbb{C}$-vector space of Bruhat-Schwartz
functions is denoted by $\mathcal{D}:=\mathcal{D}(%
\mathbb{Q}
_{p},\mathbb{C})$. We will denote by $\mathcal{D}_{\mathbb{R}}:=\mathcal{D}(%
\mathbb{Q}
_{p}^{n},\mathbb{R})$, the $\mathbb{R}$-vector space of test functions.

Given $\rho\in\lbrack0,\infty)$, we denote by $L^{\rho}:=L^{\rho}\left(
\mathbb{Q}
_{p}\right)  :=L^{\rho}\left(
\mathbb{Q}
_{p},dx\right)  ,$ the $%
\mathbb{C}
-$vector space of all the complex valued functions $g$ satisfying $\int_{%
\mathbb{Q}
_{p}}\left\vert g\left(  x\right)  \right\vert ^{\rho}dx<\infty$, and
$L^{\infty}\allowbreak:=L^{\infty}\left(
\mathbb{Q}
_{p}\right)  =L^{\infty}\left(
\mathbb{Q}
_{p},dx\right)  $ denotes the $%
\mathbb{C}
-$vector space of all the complex valued functions $g$ such that the essential
supremum of $|g|$ is bounded. The corresponding $\mathbb{R}$-vector spaces are
denoted as $L_{\mathbb{R}}^{\rho}\allowbreak:=L_{\mathbb{R}}^{\rho}\left(
\mathbb{Q}
_{p}\right)  =L_{\mathbb{R}}^{\rho}\left(
\mathbb{Q}
_{p},dx\right)  $, $1\leq\rho\leq\infty$.

\subsection{Integration on $\mathbb{Q}_{p}$}

Since $(\mathbb{Q}_{p},+)$ is a locally compact topological group, there
exists a Borel measure $dx$, called the Haar measure of $(\mathbb{Q}_{p},+)$,
unique up to multiplication by a positive constant, such that $\int_{U}dx>0$
for every non-empty Borel open set $U\subset\mathbb{Q}_{p}$, and satisfying
$\int_{E+z}dx=\int_{E}dx$ for every Borel set $E\subset\mathbb{Q}_{p}$, see
e.g. \cite[Chapter XI]{Halmos}. If we normalize this measure by the condition
$\int_{\mathbb{Z}_{p}}dx=1$, then $dx$ is unique. From now on we denote by
$dx$ the normalized Haar measure of $(\mathbb{Q}_{p},+)$.

A test function $\varphi:$ $\mathbb{Q}_{p}\rightarrow\mathbb{C}$ can be
expressed as a linear combination of characteristic functions of the form
$\varphi\left(  x\right)  =\sum_{i=1}^{l}c_{i}\Omega\left(  p^{-r_{i}}%
|x-a_{i}|_{p}\right)  $, where $c_{i}\in\mathbb{C}$ and $\Omega\left(
p^{-r_{i}}|x-a_{i}|_{p}\right)  $ is the characteristic function of $\ $the
ball $a_{i}+p^{-r_{i}}\mathbb{Z}_{p}$, for every $i$. In this case
\[%
{\displaystyle\int\limits_{\mathbb{Q}_{p}}}
\varphi\left(  x\right)  dx=\sum_{i=1}^{l}c_{i}%
{\displaystyle\int\limits_{a_{i}+p^{-r_{i}}\mathbb{Z}_{p}}}
dx=\sum_{i=1}^{l}c_{i}%
{\displaystyle\int\limits_{p^{-r_{i}}\mathbb{Z}_{p}}}
dx=\sum_{i=1}^{l}c_{i}p^{r_{i}},
\]
where we use the facts that $dx$ is invariant under translations and that
$\int_{p^{-r_{i}}\mathbb{Z}_{p}}dx=p^{r_{i}}$. By using the fact that
$\mathcal{D}\left(  \mathbb{Q}_{p}\right)  $ is a dense subspace of
$C_{0}\left(  \mathbb{Q}_{p}\right)  $, the space of continuous functions with
compact support, the functional \ $\varphi\rightarrow\int_{\mathbb{Q}_{p}%
}\varphi\left(  x\right)  dx$, $\varphi\in\mathcal{D}\left(  \mathbb{Q}%
_{p}\right)  $ has a unique extension to $C_{0}\left(  \mathbb{Q}_{p}\right)
$. For integrating more general functions, say locally integrable functions,
\ the following notion of improper integral is used.

\begin{definition}
A function $\varphi\in L_{loc}^{1}$ is said to be integrable in $\mathbb{Q}%
_{p}$ if%
\[
\lim_{m\rightarrow+\infty}%
{\displaystyle\int\limits_{B_{m}\left(  0\right)  }}
\varphi\left(  x\right)  dx=\lim_{m\rightarrow+\infty}%
{\displaystyle\sum\limits_{j=-\infty}^{m}}
\text{ }%
{\displaystyle\int\limits_{S_{j}\left(  0\right)  }}
\varphi\left(  x\right)  dx
\]
exists. If the limit exists, it is denoted as $%
{\textstyle\int\nolimits_{\mathbb{Q}_{p}}}
\varphi\left(  x\right)  dx$, and we say that the\textbf{\ }(improper)
integral exists.
\end{definition}

\subsection{Analytic change of variables}

A function $h:U\rightarrow\mathbb{Q}_{p}$ is said to be \textit{analytic} on
an open subset $U\subset\mathbb{Q}_{p}$, if for every $b\in U$ there exists an
open subset $\widetilde{U}\subset U$, with $b\in\widetilde{U}$, and a
convergent power series $\sum_{i}a_{i}\left(  x-b\right)  ^{i}$ for
$x\in\widetilde{U}$, such that $h\left(  x\right)  =\sum_{i\in\mathbb{N}}%
a_{i}\left(  x-b\right)  ^{i}$ for $x\in\widetilde{U}$. In this case,
$\frac{d}{dx}h\left(  x\right)  =\sum_{i\in\mathbb{N}}a_{i}\frac{d}{dx}\left(
x-b\right)  ^{i}$ is a convergent power series.

Let $U$, $V$ be open subsets of $\mathbb{Q}_{p}$. Let $\varphi:V$
$\rightarrow\mathbb{C}$ be a continuous function with compact support, and let
$h:U\rightarrow V$ $\ $be an analytic mapping. Then
\[%
{\textstyle\int\limits_{V}}
\varphi\left(  y\right)  dy=%
{\textstyle\int\limits_{U}}
\varphi\left(  h(x)\right)  \left\vert \frac{d}{dx}h(x)\right\vert
_{p}dx\text{,}%
\]
see e.g. \cite{V-V-Z}.

\subsection{Fourier transform}

Set $\chi_{p}(y)=\exp(2\pi i\{y\}_{p})$ for $y\in%
\mathbb{Q}
_{p}$. The map $\chi_{p}(\cdot)$ is an additive character on $%
\mathbb{Q}
_{p}$, i.e. a continuous map from $\left(
\mathbb{Q}
_{p},+\right)  $ into $S$ (the unit circle considered as multiplicative group)
satisfying $\chi_{p}(x_{0}+x_{1})=\chi_{p}(x_{0})\chi_{p}(x_{1})$,
$x_{0},x_{1}\in%
\mathbb{Q}
_{p}$. The additive characters of $%
\mathbb{Q}
_{p}$ form an Abelian group which is isomorphic to $\left(
\mathbb{Q}
_{p},+\right)  $, the isomorphism is given by $\xi\rightarrow\chi_{p}(\xi x)$,
see e.g. \cite[Section 2.3]{Alberio et al}.

If $f\in L^{1}$ its Fourier transform is defined by
\[
(\mathcal{F}f)(\xi)=\int_{%
\mathbb{Q}
_{p}}\chi_{p}(\xi x)f(x)dx,\quad\text{for }\xi\in%
\mathbb{Q}
_{p}.
\]
We will also use the notation $\mathcal{F}_{x\rightarrow\xi}f$ and
$\widehat{f}$\ for the Fourier transform of $f$. The Fourier transform is a
linear isomorphism from $\mathcal{D}$ onto itself satisfying
\begin{equation}
(\mathcal{F}(\mathcal{F}f))(\xi)=f(-\xi), \label{FF(f)}%
\end{equation}
for every $f\in\mathcal{D},$ see e.g. \cite[Section 4.8]{Alberio et al}. If
$f\in L^{2}$, its Fourier transform is defined as
\[
(\mathcal{F}f)(\xi)=\lim_{k\rightarrow\infty}\int_{|x|_{p}\leq p^{k}}\chi
_{p}(\xi\cdot x)f(x)d^{n}x,\quad\text{for }\xi\in%
\mathbb{Q}
_{p}\text{,}%
\]
where the limit is taken in $L^{2}$. We recall that the Fourier transform is
unitary on $L^{2},$ i.e. $||f||_{L^{2}}=||\mathcal{F}f||_{L^{2}}$ for $f\in
L^{2}$ and that (\ref{FF(f)}) is also valid in $L^{2}$, see e.g. \cite[Chapter
$III$, Section 2]{Taibleson}.

\section{The model}

A replicator is a model of an entity with the template property, which means
that it serves as a pattern for the generation of another replicator. This
copying process is subject to errors (mutations). Along this article we will
use replicators, genomes, macromolecules and sequences as synonyms. The
assumption of the existence of replicators implies that the information stored
in the replicators is modified randomly, and that part of it is fixed due to
the selection pressure, which in turn is related with the self-replicate
capacity of the replicators (their fitness).

In this section we introduce a $p$-adic version of Eigen's equation, see
e.g.\cite{Eigen1971}, \cite{Eigen et al}, \cite{Nowak}, \cite{Szat-PTRSB},
\cite{Tannembaum et al}, \cite{SchusterPeter}, which describes
mutation-selection process of replicating sequences.

\subsection{The space of sequences}

In our model each sequence corresponds to a $p$-adic number:%
\[
x=x_{-m}p^{-m}+x_{-m+1}p^{-m+1}+\ldots+x_{0}+x_{1}p+\ldots
\]
where the digits $x_{i}$s run through the set $\left\{  0,1,\ldots
,p-1\right\}  $. In the case $p=2$, the sequences are binary words.
Consequently, in our model the sequences are words of arbitrary length written
in the alphabet $0$, $1$,$\ldots$, $p-1$, and the space of sequences is
$\left(  \mathbb{Q}_{p},\left\vert \cdot\right\vert _{p}\right)  $, which is
an infinite set. A key feature of our model \ is the use of sequences of
variable length. It is relevant to mention that Poole et al. \cite{Poole etal}
and Scheuring \cite{Sch} already proposed solutions for Eigen's paradox which
require the usage of sequences of variable length.

On the other hand, $p$-adic numbers naturally appear in models of the genetic
code. For an in-depth discussion the reader may consult \cite{Dragovich},
\cite{KKGentic}, \cite[Chapter 4]{KKZuniga} and the references therein.

\subsection{Concentrations}

The concentration of sequence $x\in\mathbb{Q}_{p}$ at the time $t\geq0$ is
denoted as $X\left(  x,t\right)  $, this is a \ real number between zero and
one. In addition, we assume that%
\begin{equation}
\int\nolimits_{\mathbb{Q}_{p}}X\left(  y,t\right)  dy=1\text{ for }t>0\text{.}
\label{hypotesis_1}%
\end{equation}
This last condition assures that the total concentration remains constant for
$t>0$.

\subsection{\label{Section_mutation_measure}The mutation measure}

We fix a function $Q:\mathbb{R}_{+}\mathbb{\rightarrow}\left[  0,1\right]  $,
such that
\[
0\leq Q(\left\vert x\right\vert _{p})\leq1\text{, }\int\nolimits_{\mathbb{Q}%
_{p}}Q\left(  \left\vert y\right\vert _{p}\right)  dy=1.
\]
We call $Q(\left\vert x\right\vert _{p})dx$ \textit{the mutation measure}.
Given a Borel subset $E\subseteq\mathbb{Q}_{p}$ and $x\in\mathbb{Q}_{p}$, the
integral
\[
\int\nolimits_{E}Q\left(  \left\vert x-y\right\vert _{p}\right)  dy
\]
gives the probability that sequence $x$ will mutate into a sequence belonging
to $E$. In addition to this hypothesis we do not need additional assumptions
about the mutation mechanism. The results presented in this article can be
easily extended to case in which the mutation measure depends on time, more
precisely, when $Q(\left\vert x\right\vert _{p},t)dx$ is the transition
function of a Markov process.

\subsection{The\textit{ }fitness function}

The\textit{ fitness function} $f$ is a non-negative real-valued test function.
Notice that $f$ is a radial function, and by abuse of notation, we will use
the notation $f\left(  \left\vert x\right\vert _{p}\right)  $. The assumption
that function $f$ has compact support says that the evolution process is
limited to a certain region of the space of sequences, which is infinite. This
assumption allows very general fitness landscapes, for instance,%
\[
\left\{
\begin{array}
[c]{lll}%
c_{0} & \text{if} & \left\vert x\right\vert _{p}\leq p^{L_{0}}\\
g\left(  \left\vert x\right\vert _{p}\right)  & \text{if} & p^{L_{0}%
}<\left\vert x\right\vert _{p}\leq p^{L_{1}}\\
0 & \text{if} & \left\vert x\right\vert _{p}>p^{L_{1}},
\end{array}
\right.
\]
where $c_{0}$ is a nonnegative constant, $L_{0}$, $L_{1}\in\mathbb{Z}$, with
$L_{0}<$ $L_{1}$, and $g:\mathbb{R}_{+}\rightarrow\mathbb{R}_{+}$ is a function.

\subsection{The non-Archimedean replicator equation}

We set%
\[
\left(  \boldsymbol{W}\varphi\right)  \left(  x\right)  =Q\left(  \left\vert
x\right\vert _{p}\right)  \ast\left\{  f\left(  \left\vert x\right\vert
_{p}\right)  \varphi\left(  x\right)  \right\}  =\int\nolimits_{\mathbb{Q}%
_{p}}Q\left(  \left\vert x-y\right\vert _{p}\right)  f\left(  \left\vert
y\right\vert _{p}\right)  \varphi\left(  y\right)  dy,
\]
where $Q$ and $f$ are as before. Notice that for $1\leq\rho\leq\infty$,
\[%
\begin{array}
[c]{ccc}%
L^{\rho}(\mathbb{Q}_{p},\mathbb{C}) & \rightarrow & L^{\rho}(\mathbb{Q}%
_{p},\mathbb{C})\\
&  & \\
\varphi & \rightarrow & \boldsymbol{W}\varphi
\end{array}
\]
is a well-defined continuous operator.

Now our non-Archimedean mutation-selection equation has the form%

\begin{equation}
\frac{\partial X\left(  x,t\right)  }{\partial t}=\boldsymbol{W}X\left(
x,t\right)  -\Phi\left(  t\right)  X\left(  x,t\right)  \text{, }%
x\in\mathbb{Q}_{p}\text{, }t\in\mathbb{R}_{+}\text{,} \label{eq1}%
\end{equation}
where%
\[
\Phi\left(  t,X\right)  :=\Phi\left(  t\right)  =\int\nolimits_{\mathbb{Q}%
_{p}}f\left(  \left\vert y\right\vert _{p}\right)  X\left(  y,t\right)
dy\text{ \ \ for }t\geq0\text{.}%
\]
This function $\Phi\left(  t\right)  $\ is used to maintain constant the total
concentration in the chemostat. Later on, we will consider the Cauchy problem
associated to (\ref{eq1}) with initial datum%
\begin{equation}
X\left(  x,0\right)  =X_{0}\in\mathcal{D}_{\mathbb{R}} \label{eq1_A}%
\end{equation}

By changing variables as
\begin{equation}
X\left(  x,t\right)  =Y\left(  x,t\right)  \exp\left(  -\int_{0}^{t}%
\Phi\left(  \tau\right)  d\tau\right)  , \label{eq1A}%
\end{equation}
like in the classical case, (\ref{eq1})-(\ref{eq1_A}) becomes%
\begin{equation}
\left\{
\begin{array}
[c]{lll}%
\frac{\partial Y\left(  x,t\right)  }{\partial t}=\boldsymbol{W}Y\left(
x,t\right)  , & x\in\mathbb{Q}_{p}, & t\in\mathbb{R}_{+}\\
&  & \\
Y\left(  x,0\right)  =X_{0}\in\mathcal{D}_{\mathbb{R}}. &  &
\end{array}
\right.  \label{eq2}%
\end{equation}

\subsection{Comments}

(i) Archimedean evolution equations appear in several models of evolution, see
e.g. \cite{Perthame} and \cite{Andre-Vaillo et al}. In particular the proposed
equation (\ref{eq1}) makes sense in $\mathbb{R}$, however, the interpretation
of the space variable $x$ as a sequence of variable length is only natural in
the $p$-adic setting.

(ii) The Equation (\ref{eq1}) is really a family of models depending on $f$,
$Q$. We formulated these models in dimension one, that is, just for `one type
of sequences'. However, the equations introduced here can be extended to
dimension $n$, that is, to the case of $n$ `types of sequences'. In principle,
the models introduced here may be extended to include recombination and sexual reproduction.

\section{\label{SEct_P_adic_replicator_Eq}An ultrametric Version of the
classical replicator equation}

Eigen's evolution model consists of a systems of non-linear ordinary
differential equations describing the evolution of the concentrations of
sequences, which belong to a finite metric space, where the distance is
induced by the Hamming weight. If we use binary sequences, the space of
sequences is $\mathbb{Z}_{2}/2^{2M}\mathbb{Z}_{2}$, for some positive integer
$M$.

In this section by using ideas from \cite{zuniga-Nonlieal}, we show a
discretization of the equation (\ref{eq1}) that agrees with the Eigen model in
an ultrametric space formed by finite $p$-adic sequences, where the mutation
probability is a function of the ultrametric distance between sequences. By a
suitable normalization the space of sequences is $\mathbb{Z}_{p}%
/p^{2M}\mathbb{Z}_{p}$, and the ultrametric distance is induced by the
restriction of the $p$-adic norm to $\mathbb{Z}_{p}/p^{2M}\mathbb{Z}_{p}$.
This fact shows that our equation (\ref{eq1}) is a non-Archimedean
generalization of the Eigen model.

\begin{notation}
The set of non-negative integers is denoted as $\mathbb{N}$.
\end{notation}

\subsection{The space of sequences $G_{M}$}

We fix $M\in\mathbb{N\smallsetminus}\left\{  0\right\}  $ and set
\[
G_{M}:=p^{-M}\mathbb{Z}_{p}/p^{M}\mathbb{Z}_{p}.
\]
We consider $G_{M}$\ as an additive group and fix the following systems of
representatives: any $I\in G_{M}$ is represented as%
\begin{equation}
I=I_{-M}p^{-M}+I_{-M+1}p^{-M+1}+\cdots+I_{0}+\cdots+I_{M-1}p^{M-1},
\label{seq_I}%
\end{equation}
where the $I_{j}$s belong to $\left\{  0,1,\ldots,p-1\right\}  $. Furthermore,
the restriction of $\left\vert \cdot\right\vert _{p}$ to $G_{M}$ induces an
absolute value such that $\left\vert G_{M}\right\vert _{p}=\left\{
0,p^{-\left(  M+1\right)  },\cdots,p^{-1},1,\cdots,p^{M}\right\}  $. We endow
$G_{M}$ with the metric induced by $\left\vert \cdot\right\vert _{p}$, and
thus $G_{M}$ becomes a finite ultrametric space. In addition, $G_{M}$ can be
identified with the set of branches (vertices at the top level) of a rooted
tree with $2M+1$ levels and $p^{2M}$ branches. Any element $I\in G_{M}$ can be
uniquely written as $p^{M}\widetilde{I}$, where%
\[
\widetilde{I}=\widetilde{I}_{0}+\widetilde{I}_{1}p+\cdots+\widetilde{I}%
_{2M-1}p^{2M-1}\in\mathbb{Z}_{p}/p^{2M}\mathbb{Z}_{p},
\]
with the $\widetilde{I}_{j}$s belonging to $\left\{  0,1,\ldots,p-1\right\}
$. The elements of the $\mathbb{Z}_{p}/p^{2M}\mathbb{Z}_{p}$ are in bijection
with the vertices at the top level of the above mentioned rooted tree. By
definition the root of the tree is the only vertex at level $0$. There are
exactly $p$ vertices at level $1$, which correspond with the possible values
of the digit $\widetilde{I}_{0}$ in the $p$-adic expansion of $\widetilde{I}$.
Each of these vertices is connected to the root by a non-directed edge. At
level $\ell$, with $1\leq\ell\leq2M$, there are exactly $p^{\ell}$ vertices,
\ each vertex corresponds to a truncated expansion of $\widetilde{I}$ of the
form $\widetilde{I}_{0}+\cdots+\widetilde{I}_{\ell-1}p^{\ell-1}$. The vertex
corresponding to $\widetilde{I}_{0}+\cdots+\widetilde{I}_{\ell-1}p^{\ell-1}$
is connected to a vertex $\widetilde{I}_{0}^{\prime}+\cdots+\widetilde
{I}_{\ell-2}^{\prime}p^{\ell-2}$ at the level $\ell-1$ if and only if $\left(
\widetilde{I}_{0}+\cdots+\widetilde{I}_{\ell-1}p^{\ell-1}\right)  -\left(
\widetilde{I}_{0}^{\prime}+\cdots+\widetilde{I}_{\ell-2}^{\prime}p^{\ell
-2}\right)  $ is divisible by $p^{\ell-1}$. Notice that there are other
geometric realizations of $G_{M}$. For instance, we can chose a forest formed
by $p$ rooted trees, each of them with $2M$ levels and $p^{2M-1}$ branches.

\subsection{\label{Sect_DIs}A discretization of equation (\ref{eq1})}

We denote by $\mathcal{D}_{M}^{-M}$ the $\mathbb{R}$-vector subspace of
$\mathcal{D}_{\mathbb{R}}$ spanned by the functions%
\[
\Omega\left(  p^{M}\left\vert x-I\right\vert _{p}\right)  ,\text{ }I\in
G_{M}\text{.}%
\]
Notice that $\Omega\left(  p^{M}\left\vert x-I\right\vert \right)
\Omega\left(  p^{M}\left\vert x-J\right\vert \right)  =0$ for any $x$, if
$I\neq J$. Thus, any function $\varphi\in\mathcal{D}_{M}^{-M}$ has the form%
\[
\varphi\left(  x\right)  =\sum_{I\in G_{M}}\varphi\left(  I\right)
\Omega\left(  p^{M}\left\vert x-I\right\vert _{p}\right)  ,
\]
where the $\varphi\left(  I\right)  $s are real numbers. Notice that the
dimension of $\mathcal{D}_{M}^{-M}\left(  \mathbb{Q}_{p}\right)  $ is
$\#G_{M}=p^{2M}$.

In order to explain the connection between the non-Archimedean replicator
equation (\ref{eq1}) and the classical one, we assume that $Q\left(
\left\vert x\right\vert _{p}\right)  $, $f\left(  \left\vert x\right\vert
_{p}\right)  $ and $X(x,t)$ belong to $\mathcal{D}_{M}^{-M}$, for any $t$.
Then%
\[
Q\left(  \left\vert x\right\vert _{p}\right)  =\frac{1}{C_{M}}\sum_{I\in
G_{M}}Q\left(  \left\vert I\right\vert _{p}\right)  \Omega\left(
p^{M}\left\vert x-I\right\vert _{p}\right)  ,
\]
with $C_{M}=p^{-M}\sum_{I\in G_{M}}Q\left(  \left\vert I\right\vert
_{p}\right)  $,
\[
f\left(  \left\vert x\right\vert _{p}\right)  =\sum_{I\in G_{M}}f\left(
\left\vert I\right\vert _{p}\right)  \Omega\left(  p^{M}\left\vert
x-I\right\vert _{p}\right)  ,
\]
and%
\[
X(x,t)=\sum_{I\in G_{M}}X(I,t)\Omega\left(  p^{M}\left\vert x-I\right\vert
_{p}\right)  \text{ for any }t\geq0\text{,}%
\]
where each $X(I,t)$ is a real-valued function of class $C^{1}$ in $t$. Now%
\begin{gather}
\int\nolimits_{\mathbb{Q}_{p}}Q\left(  \left\vert x-y\right\vert _{p}\right)
f\left(  \left\vert y\right\vert _{p}\right)  X\left(  y,t\right)
dy=\label{eq3}\\
\left\{  \frac{1}{C_{M}}\sum_{K\in G_{M}}\sum_{I\in G_{M}}Q\left(  \left\vert
K\right\vert _{p}\right)  f\left(  \left\vert I\right\vert _{p}\right)
X(I,t)\right\}  \Omega\left(  p^{M}\left\vert x-K\right\vert _{p}\right)
\ast\Omega\left(  p^{M}\left\vert x-I\right\vert _{p}\right)  ,\nonumber
\end{gather}
and by using that $\Omega\left(  p^{M}\left\vert x-K\right\vert _{p}\right)
\ast\Omega\left(  p^{M}\left\vert x-I\right\vert _{p}\right)  =p^{M}%
\Omega\left(  p^{M}\left\vert x-\left(  I+K\right)  \right\vert _{p}\right)  $
and the fact that $G_{M}$ \ is an additive group, the right-hand side of
(\ref{eq3}) can be rewritten as
\[
\left\{  \frac{1}{C}\sum_{I\in G_{M}}Q\left(  \left\vert J-I\right\vert
_{p}\right)  f\left(  \left\vert I\right\vert _{p}\right)  X(I,t)\right\}
\Omega\left(  p^{M}\left\vert x-J\right\vert _{p}\right)  ,
\]
with $C=\sum_{I\in G_{M}}Q\left(  \left\vert I\right\vert _{p}\right)  $.
Finally, using the fact that the $\Omega\left(  p^{M}\left\vert x-I\right\vert
_{p}\right)  $, $I\in G_{M}$ are $\mathbb{R}$-linearly independent, we get%
\begin{equation}
\frac{d}{dt}X\left(  J,t\right)  =\frac{1}{C}\sum_{I\in G_{M}}Q\left(
\left\vert J-I\right\vert _{p}\right)  f\left(  \left\vert I\right\vert
_{p}\right)  X(I,t)-\Phi_{M}\left(  t\right)  X\left(  J,t\right)  \text{
\ for }J\in G_{M}, \label{eq_Eigen}%
\end{equation}
where
\[
\Phi_{M}\left(  t\right)  =p^{-M}\sum_{I\in G_{M}}f\left(  \left\vert
I\right\vert _{p}\right)  X(I,t),
\]
which is exactly the Eigen model on $G_{M}$, but $G_{M}$ is an ultrametric
space, where the distance comes from the $p$-adic norm.

\subsection{The limit $M$ tends to infinity}

Heuristically speaking, the limit of the system of equations (\ref{eq_Eigen})
when $M$ tends to infinity is the evolution equation (\ref{eq1}). In
\cite{Av-8},\ Avetisov, Kozyrev et al. established (using physical arguments)
that certain non-Archimedean kinetic models, similar to (\ref{eq_Eigen}), have
`continuous $p$-adic limits' as reaction-ultradiffusion equations. A
mathematical explanation of these constructions is given in our article
\cite{zuniga-Nonlieal}. From a mathematical perspective, the Cauchy problem
associated with (\ref{eq1}) can be very well approximated by an initial value
problem associated with the system (\ref{eq_Eigen}), in the sense that any
solution of (\ref{eq1}) is arbitrarily close to a suitable solution of
(\ref{eq_Eigen}) in a certain function space, when $M$ tends to infinity, see
introduction of \cite{zuniga-Nonlieal}. This result follows by applying
techniques used in \cite{zuniga-Nonlieal} to the system (\ref{eq2}) with
$X_{0}$ a test function, see (\ref{eq2}).

Now, if consider the system (\ref{eq_Eigen}) as a system of ODEs in
$\mathbb{R}^{p^{2M}}$, then it seems not plausible that the `limit
$N\rightarrow\infty$' can be defined.

\section{\label{Section-Maynard-Smith}The Maynard Smith Ansatz}

In this section we present a $p$-adic version of Maynard Smith approach to the
error threshold problem, see \cite{Maynad Smith}, \cite{Szat-PTRSB}. We divide
the space of sequences into two disjoint sets:
\begin{equation}
\mathbb{Q}_{p}=\left[  I+p^{M}\mathbb{Z}_{p}\right]  \bigsqcup\left[
\mathbb{Q}_{p}\smallsetminus I+p^{M}\mathbb{Z}_{p}\right]  ,
\label{condition_0}%
\end{equation}
where $I$ is a fixed sequence as in (\ref{seq_I}), with $M\geq1$, and assume
that
\begin{equation}
f\mid_{I+p^{M}\mathbb{Z}_{p}}\equiv a\text{, \ \ }f\mid_{\mathbb{Q}%
_{p}\smallsetminus I+p^{M}\mathbb{Z}_{p}}\equiv b\text{, \ with }a>b\text{,}
\label{condition_1}%
\end{equation}
here \textquotedblleft$\equiv$\textquotedblright\ means identically equal.
This means that the group $I+p^{M}\mathbb{Z}_{p}$ contains the fittest
sequences and that all these sequences coincide with $I$ up to digit $I_{M-1}%
$. We denote by $X(x,t)$ the concentration of $I+p^{M}\mathbb{Z}_{p}$ and by
$Y(x,t)$ the concentration of $\mathbb{Q}_{p}\smallsetminus\left[
I+p^{M}\mathbb{Z}_{p}\right]  $. Notice that the supports of $X(x,t)$\ and
$Y(x,t)$\ are disjoint. We denote by $q:=q\left(  M,Q\right)  $ the
probability that a sequence in $I+p^{M}\mathbb{Z}_{p}$ mutates into a sequence
belonging to $\mathbb{Q}_{p}\smallsetminus\left[  I+p^{M}\mathbb{Z}%
_{p}\right]  $, and we denote by $r:=r\left(  M,Q\right)  $ the probability of
mutation of a sequence from $\mathbb{Q}_{p}\smallsetminus\left[
I+p^{M}\mathbb{Z}_{p}\right]  $ into a sequence in $I+p^{M}\mathbb{Z}_{p}$.
The system of equations governing the development of these populations is
\begin{equation}%
\begin{array}
[c]{cc}%
\frac{\partial X(x,t)}{\partial t}= & a\left(  1-q\right)
X(x,t)+brY(x,t)-\Phi\left(  t\right)  X\left(  x,t\right) \\
& \\
\frac{\partial Y(x,t)}{\partial t}= & aqX(x,t)+b\left(  1-r\right)
Y(x,t)-\Phi\left(  t\right)  Y\left(  x,t\right)  ,
\end{array}
\label{system1}%
\end{equation}
where
\[
\int\limits_{\mathbb{Q}_{p}}X\left(  x,t\right)  dx+\int\limits_{\mathbb{Q}%
_{p}}Y\left(  x,t\right)  dx=1,
\]
and
\begin{align*}
\Phi\left(  t\right)   &  =\int\limits_{I+p^{M}\mathbb{Z}_{p}}f\left(
\left\vert x\right\vert _{p}\right)  X\left(  x,t\right)  dx+\int
\limits_{\mathbb{Q}_{p}\smallsetminus I+p^{M}\mathbb{Z}_{p}}f\left(
\left\vert x\right\vert _{p}\right)  Y\left(  x,t\right)  dx\\
&  =a\int\limits_{I+p^{M}\mathbb{Z}_{p}}X\left(  x,t\right)  dx+b\int
\limits_{\mathbb{Q}_{p}\smallsetminus I+p^{M}\mathbb{Z}_{p}}Y\left(
x,t\right)  dx.
\end{align*}
We assume that for $M$ sufficiently large,\ $r\left(  M,Q\right)  $ is very
small (this condition should be verified for each particular choice of $Q$),
so we can assume that system (\ref{system1}) has the form%
\begin{equation}%
\begin{array}
[c]{cc}%
\frac{\partial X(x,t)}{\partial t}= & a\left(  1-q\right)  X(x,t)-\Phi\left(
t\right)  X\left(  x,t\right) \\
& \\
\frac{\partial Y(x,t)}{\partial t}= & aqX(x,t)+bY(x,t)-\Phi\left(  t\right)
Y\left(  x,t\right)  .
\end{array}
\label{system2}%
\end{equation}
By taking $Z(x,t)=\frac{X(x,t)}{Y(x,t)}$, system (\ref{system2}) becomes%
\[
\frac{\partial Z(x,t)}{\partial t}=Z(x,t)\left\{  a\left(  1-q\right)
-aqZ(x,t)-b\right\}  .
\]
Assuming that concentration $Z(x,t)$ achieves a steady concentration
$\overline{Z}(x)$ over the time, we get%
\[
\overline{Z}(x)=\frac{a\left(  1-q\right)  -b}{aq}.
\]
The original population persists, i.e. the sequences in $I+p^{M}\mathbb{Z}%
_{p}$ survive in a long term, if and only if $\overline{Z}(x)>0$, i.e. if and
only if
\[
1-q>\frac{b}{a}.
\]
By writing $\frac{b}{a}=1-s$, with $s\in\left(  0,1\right)  $, \ the error
threshold \ is given by%
\begin{equation}
q<s. \label{error_thr}%
\end{equation}
This is exactly the classical condition determining the error threshold, see
e.g. \cite{Maynad Smith}, \cite{Szat-PTRSB}.

\subsection{Comments}

Notice that by using the ultrametric property $\left\vert x-y\right\vert
_{p}=\left\vert y\right\vert _{p}$ \ for $x\in p^{M}\mathbb{Z}_{p}$ and
$y\in\mathbb{Q}_{p}\smallsetminus p^{M}\mathbb{Z}_{p}$, we have
\begin{align*}
q  &  =q\left(  M,\alpha\right)  =\int\limits_{I+p^{M}\mathbb{Z}_{p}}\text{
}\int\limits_{\mathbb{Q}_{p}\smallsetminus\left[  I+p^{M}\mathbb{Z}%
_{p}\right]  }Q\left(  \left\vert x-y\right\vert _{p}\right)  dydx\\
&  =\int\limits_{p^{M}\mathbb{Z}_{p}}\text{ }\int\limits_{\mathbb{Q}%
_{p}\smallsetminus p^{M}\mathbb{Z}_{p}}Q\left(  \left\vert x-y\right\vert
_{p}\right)  dydx=\int\limits_{p^{M}\mathbb{Z}_{p}}\text{ }\int
\limits_{\mathbb{Q}_{p}\smallsetminus p^{M}\mathbb{Z}_{p}}Q\left(  \left\vert
y\right\vert _{p}\right)  dydx\\
&  =p^{-M}\int\limits_{\mathbb{Q}_{p}\smallsetminus p^{M}\mathbb{Z}_{p}%
}Q\left(  \left\vert y\right\vert _{p}\right)  dy=p^{-M}\int
\limits_{\text{supp }Q\cap\left[  \mathbb{Q}_{p}\smallsetminus p^{M}%
\mathbb{Z}_{p}\right]  }Q\left(  \left\vert y\right\vert _{p}\right)  dy,
\end{align*}
which implies that $q$ is independent of $I$, $r=r\left(  M,Q\right)  =q$, and
that conditions (\ref{condition_0})-(\ref{condition_1}) can be replaced by
\[
\text{supp }Q=\left[  I+p^{M}\mathbb{Z}_{p}\right]  \bigsqcup\left[
\text{supp }Q\smallsetminus\left[  I+p^{M}\mathbb{Z}_{p}\right]  \right]  ,
\]
where $I$ is a fixed sequence as in (\ref{seq_I}), with $M\geq1$, and by
\[
f\mid_{I+p^{M}\mathbb{Z}_{p}}\equiv a\text{, \ \ }f\mid_{\text{supp
}Q\smallsetminus\left[  I+p^{M}\mathbb{Z}_{p}\right]  }\equiv b\text{,
\ \ with }a>b\text{.}%
\]

\section{\label{Section_Error_Eigen_p_adic_model}The Error Threshold Problem
Using a mutation measure supported in unit ball}

In this section we study the error threshold problem using the Maynard Smith
ansatz with a mutation measure supported in the unit ball.

\subsection{A class of mutation measures supported in the unit ball}

Take $\alpha\geq0$, and consider
\begin{equation}
Q\left(  \left\vert x\right\vert _{p};\alpha\right)  =\frac{\left\vert
x\right\vert _{p}^{\alpha}\Omega\left(  \left\vert x\right\vert _{p}\right)
}{Z\left(  \alpha\right)  }, \label{EC_1}%
\end{equation}
where for $\gamma\in\mathbb{R}$,
\begin{equation}
Z\left(  \gamma\right)  =\int_{\mathbb{Z}_{p}}\left\vert x\right\vert
_{p}^{\gamma}dx=\frac{1-p^{-1}}{1-p^{-1-\gamma}}\text{ for }\gamma>-1.
\label{EC_2}%
\end{equation}

Then, $Q\left(  \left\vert x\right\vert _{p};\alpha\right)  dx$ gives rise to
a family of mutation measures, which include the uniform distribution for
$\alpha=0$.

We now fix a sequence $I\in\mathbb{Z}_{p}$, which plays the role of the master
sequence, and divide the space of sequences $\mathbb{Z}_{p}$ into two subsets:
$I+p^{M}\mathbb{Z}_{p}$ and $\mathbb{Z}_{p}\smallsetminus\left[
I+p^{M}\mathbb{Z}_{p}\right]  $ for some positive integer $M$. The set
$I+p^{M}\mathbb{Z}_{p}$ consists of the sequences in the unit ball that
coincide with the sequence $I$ up to the digit $M-1$. We denote by $H_{M}$ a
fixed set of representatives of $\mathbb{Z}_{p}/p^{M}\mathbb{Z}_{p}$. We also
assume that%
\[
f\mid_{I+p^{M}\mathbb{Z}_{p}}\equiv a\text{, }f\mid_{\mathbb{Z}_{p}%
\smallsetminus I+p^{M}\mathbb{Z}_{p}}\equiv b\text{, with }a>b\text{.}%
\]
Notice that by the remarks made at the end of Section
\ref{Section-Maynard-Smith}, we can apply the Maynard Smith ansatz to
establish (\ref{error_thr}) under the above mentioned conditions. We now
compute the probability that a sequence in the set $I+p^{M}\mathbb{Z}_{p}$
mutates into a sequence belonging to the set $\mathbb{Z}_{p}\smallsetminus
\left[  I+p^{M}\mathbb{Z}_{p}\right]  $ (notice that this probability does not
depended on $I$):%
\begin{gather*}
q\left(  \alpha\right)  :=\frac{1}{Z\left(  \alpha\right)  }\int
\limits_{I+p^{M}\mathbb{Z}_{p}}\text{ }\int\limits_{\mathbb{Z}_{p}%
\smallsetminus I+p^{M}\mathbb{Z}_{p}}\text{ }\left\vert x-y\right\vert
_{p}^{\alpha}dydx\\
=\frac{1}{Z\left(  \alpha\right)  }\int\limits_{\left\vert x-I\right\vert
_{p}\leq p^{-M}}\text{ }\int\limits_{\left\vert y-I\right\vert _{p}>p^{-M}%
}\text{ }\left\vert x-y\right\vert _{p}^{\alpha}dydx\\
=\frac{1}{Z\left(  \alpha\right)  }\int\limits_{p^{M}\mathbb{Z}_{p}}\text{
}\int\limits_{\mathbb{Z}_{p}\smallsetminus p^{M}\mathbb{Z}_{p}}\text{
}\left\vert x-y\right\vert _{p}^{\alpha}dydx=\frac{1}{Z\left(  \alpha\right)
}\int\limits_{p^{M}\mathbb{Z}_{p}}\text{ }\int\limits_{\mathbb{Z}%
_{p}\smallsetminus p^{M}\mathbb{Z}_{p}}\text{ }\left\vert y\right\vert
_{p}^{\alpha}dydx\\
=\frac{p^{-M}}{Z\left(  \alpha\right)  }\int\limits_{\mathbb{Z}_{p}%
\smallsetminus p^{M}\mathbb{Z}_{p}}\text{ }\left\vert y\right\vert
_{p}^{\alpha}dy=\frac{p^{-2M}}{Z\left(  \alpha\right)  }\sum\limits_{J\in
H_{M}\text{, }J\neq0}\text{ }\left\vert J\right\vert _{p}^{\alpha},
\end{gather*}
where we use the ultrametric property: $\left\vert x-y\right\vert
_{p}=\left\vert y\right\vert _{p}$ for $\left\vert x\right\vert _{p}\leq
p^{-M}$ and $\left\vert y\right\vert _{p}>p^{-M}$.

Notice that%
\[
q\left(  \alpha\right)  >\frac{p^{-2M}}{Z(\alpha)}\left\vert p^{M-1}%
\right\vert _{p}^{\alpha}=\frac{p^{-2M-(M-1)\alpha}}{Z(\alpha)}>\frac
{p^{-2M-M\alpha}}{Z(\alpha)}>p^{-2M-M\alpha},
\]
since $\frac{1}{Z(\alpha)}\in\left[  1,\frac{1}{1-p^{-1}}\right)  $ for
$\alpha\in\left[  0,+\infty\right)  $.

\subsubsection{Classical error threshold: $M$ and $\alpha$ fixed}

We analyze now weather or not the condition (\ref{error_thr}) is satisfied,
when $M$ and $\alpha\in\left[  0,+\infty\right)  $ are fixed. The condition
$M$ fixed can be relaxed to `$M$ is upper bounded.' Taking into account that
$s>0$ can be arbitrarily close to zero, then there exists $M_{c}$ such that
\[
q\left(  \alpha\right)  >p^{-2M_{c}-M_{c}\alpha}\geq s,
\]
which implies the existence of a classical error threshold:%
\begin{equation}
M_{c}\leq-\frac{\ln s}{\left(  2+\alpha\right)  \ln p}\text{ for }s\in\left(
0,1\right)  \text{ and }\alpha\text{ fixed.} \label{EC_5}%
\end{equation}

\subsubsection{\label{Section_Escaping_Error_P}Overcoming Eigen's paradox: $M$
\ variable and $\alpha$ fixed}

If $M$ can grow and $\alpha$\ is fixed, the condition (\ref{error_thr}) is
satisfied if $\ p^{-2M-M\alpha}<q\left(  \alpha\right)  <s$, which implies
that%
\[
M>\frac{-\ln s}{\left(  2+\alpha\right)  \ln p}\text{ for }s\in\left(
0,1\right)  \text{.}%
\]

Under a `fierce competition'\ between the groups $I+p^{M}\mathbb{Z}_{p}$,
$\mathbb{Z}_{p}\smallsetminus\left[  I+p^{M}\mathbb{Z}_{p}\right]  $, i.e.
when rate $b$ approaches from the left to rate $a$ (i.e. $s\rightarrow0^{+}$),
$M$ must grow, which means that the survival of the sequences in the group
$I+p^{M}\mathbb{Z}_{p}$ demands that they get closer to master sequence $I$,
which means, that they must increase their lenghts. Then in this model \ the
`classical Eigen's paradox does not occur' because the length of the genomes
can grow during the evolution process.

\section{The Error Threshold Problem Using \ a mutation measure of Gibbs type}

In this section we study the error threshold problem using the Maynard Smith
ansatz with a mutation measure of Gibbs type depending only on $\left\vert
\cdot\right\vert _{p}$.

\subsection{\label{Section_Gibbs_measure}A mutation measure of Gibbs type}

We propose a mutation measure of type%
\begin{equation}
\frac{e^{-\beta E\left(  \left\vert x\right\vert _{p}\right)  }}{Z\left(
\beta,E\right)  }, \label{EC_Q_form}%
\end{equation}
where $\beta>0$, $E:\mathbb{R}_{+}\rightarrow\mathbb{R}_{+}$, and $Z\left(
\beta,E\right)  =\int_{\mathbb{Q}_{p}}e^{-\beta E\left(  \left\vert
x\right\vert _{p}\right)  }dx<\infty$. A Gibbs measure is a natural choice
when dealing with infinite systems. On the other hand, mutations matrices of
type $Q(I,J)=\frac{e^{-\beta E\left(  I,J\right)  }}{C}$ appear naturally in
the models of evolution using spin glasses technique, see e.g. \cite[Equations
\ (6) and (8)]{Let}. We pick $E\left(  \left\vert x\right\vert _{p}\right)
=\left\vert x\right\vert _{p}^{\alpha}$, with $\alpha>0$, which corresponds to
the simplest energy function. By assuming that $\beta>0$ is sufficiently large
and taking into account the fast decay of the function $\exp(-\beta\cdot)$,
our hypothesis on the mutation measure implies that the most probable
mutations of a given sequence $I$ happen to sequences which are very close to
$I$ in the $p$-adic norm, which are sequences belonging to a ball of type
$I+p^{M}\mathbb{Z}_{p}$, with $M$ sufficiently large. In practical terms, this
means that the replicators are not too dispersed on $\mathbb{Q}_{p}$. It is
interesting to quote here that according to \cite{Szabo et al}:
\textquotedblleft in silico simulations reveal that replicators with limited
dispersal evolve towards higher efficiency and fidelity.\textquotedblright

We assume that the probability that a sequence $x\in\mathbb{Q}_{p}$ mutates
into a sequence belonging to set $B$ (a Borel subset of $\mathbb{Q}_{p}$) is
given by%
\[
P(x,B;\alpha,\beta)=\frac{1}{C}\int_{B}e^{-\beta\left\vert x-y\right\vert
_{p}^{\alpha}}dy,
\]
where $C\left(  \alpha,\beta\right)  :=C$, and $\alpha$, $\beta$ are positive
constants such that $\frac{1}{C}\int_{\mathbb{Q}_{p}}e^{-\beta\left\vert
x-y\right\vert _{p}^{\alpha}}dy=1$. Notice that $P(x,B;\alpha,\beta)$ is a
space homogeneous Markov transition function, the parameter $\beta$ (which is
typically interpreted as proportional to the inverse of the temperature) plays
the role of time. We fix a sequence%
\[
I=\sum_{l=-k}^{\infty}I_{l}p^{l}\text{,}%
\]
and assume that $I$ has `infinite length', and consider the ball
$I+p^{M}\mathbb{Z}_{p}$, $M\in\mathbb{N}$, which contains all the sequences
that coincide with $I$ up to the digit $I_{M-1}$. We can consider\ $I$ as the
master sequence, and $M$ is the minimum number of nucleotides that a sequence
in the set $I+p^{M}\mathbb{Z}_{p}$ shares with $I$.

The probability that a sequence $x$ mutates into a sequence belonging to
$\mathbb{Q}_{p}\smallsetminus\left[  I+p^{M}\mathbb{Z}_{p}\right]  $ is
$P(x,\mathbb{Q}_{p}\smallsetminus\left[  I+p^{M}\mathbb{Z}_{p}\right]
;\alpha,\beta)$, and the probability that any sequence from the ball
$I+p^{M}\mathbb{Z}_{p}$ mutates into a sequence in $\mathbb{Q}_{p}%
\smallsetminus\left[  I+p^{M}\mathbb{Z}_{p}\right]  $ is given by%
\[
q\left(  M,\alpha,\beta\right)  :=\frac{1}{C}\int\limits_{I+p^{M}%
\mathbb{Z}_{p}}\text{ }\int\limits_{\mathbb{Q}_{p}\smallsetminus\left[
I+p^{M}\mathbb{Z}_{p}\right]  }e^{-\beta\left\vert x-y\right\vert _{p}%
^{\alpha}}dydx.
\]
We assert that%
\begin{equation}
q\left(  M,\alpha,\beta\right)  >\frac{\left(  p-1\right)  p^{-2M}}%
{C}e^{-\beta p^{\alpha}}. \label{q_definition}%
\end{equation}
Indeed, by using the fact that the measure $dydx$ is invariant under
translations, and the ultrametric property of $\left\vert \cdot\right\vert
_{p}$, we have
\begin{gather*}
q\left(  M,\alpha,\beta\right)  =\frac{1}{C}\int\limits_{I+p^{M}\mathbb{Z}%
_{p}}\text{ }\int\limits_{\mathbb{Q}_{p}\smallsetminus\left[  I+p^{M}%
\mathbb{Z}_{p}\right]  }e^{-\beta\left\vert \left(  x-I\right)  -\left(
y-I\right)  \right\vert _{p}^{\alpha}}dydx\\
=\frac{1}{C}\int\limits_{p^{M}\mathbb{Z}_{p}}\text{ }\int\limits_{\mathbb{Q}%
_{p}\smallsetminus p^{M}\mathbb{Z}_{p}}e^{-\beta\left\vert x-y\right\vert
_{p}^{\alpha}}dydx=\frac{1}{C}\int\limits_{\left\vert x\right\vert _{p}\leq
p^{-M}}\text{ }\int\limits_{\left\vert y\right\vert _{p}>p^{-M}}%
e^{-\beta\left\vert x-y\right\vert _{p}^{\alpha}}dydx\\
=\frac{1}{C}\int\limits_{\left\vert x\right\vert _{p}\leq p^{-M}}\text{ }%
\int\limits_{\left\vert y\right\vert _{p}>p^{-M}}e^{-\beta\left\vert
y\right\vert _{p}^{\alpha}}dydx=\frac{p^{-M}}{C}\int\limits_{\left\vert
y\right\vert _{p}>p^{-M}}e^{-\beta\left\vert y\right\vert _{p}^{\alpha}}dy\\
=\frac{p^{-M}}{C}\sum\limits_{j=-M+1}^{\infty}\int\limits_{\left\vert
y\right\vert _{p}=p^{j}}e^{-\beta\left\vert y\right\vert _{p}^{\alpha}}dy\\
>\frac{p^{-M}}{C}\int\limits_{\left\vert y\right\vert _{p}=p^{-M+1}}%
e^{-\beta\left\vert y\right\vert _{p}^{\alpha}}dy=\frac{\left(  1-p^{-1}%
\right)  p^{-2M+1}}{C}e^{-\beta p^{\left(  -M+1\right)  \alpha}}\\
=\frac{\left(  p-1\right)  p^{-2M}}{C}e^{-\beta p^{\alpha}p^{-M\alpha}}%
>\frac{\left(  p-1\right)  p^{-2M}}{C}\inf_{M\in\mathbb{N}}e^{-\beta
p^{\alpha}p^{-M\alpha}}=\frac{\left(  p-1\right)  p^{-2M}}{C}e^{-\beta
p^{\alpha}}.
\end{gather*}
We also notice that
\begin{equation}
q\left(  M,\alpha,\beta\right)  =\frac{p^{-M}}{C}\int\limits_{\left\vert
y\right\vert _{p}>p^{-M}}e^{-\beta\left\vert y\right\vert _{p}^{\alpha}%
}dy<\frac{p^{-M}}{C}\int\limits_{\mathbb{Q}_{p}}e^{-\beta\left\vert
y\right\vert _{p}^{\alpha}}dy=C_{0}p^{-M}, \label{decayq}%
\end{equation}
where $C_{0}=C_{0}(\alpha,\beta)$ is a positive constant independent of $M$,
since $e^{-\beta\left\vert y\right\vert _{p}^{\alpha}}$ is an integrable
function, see e.g. the proof of Lemma 4.1 in \cite{Koch}.

On the other hand, we denote by $r:=r\left(  M,\alpha,\beta\right)  $ the
probability of mutation of a sequence from $\mathbb{Q}_{p}\smallsetminus
\left[  I+p^{M}\mathbb{Z}_{p}\right]  $ into $I+p^{M}\mathbb{Z}_{p}$. Then
\begin{equation}
r\left(  M,\alpha,\beta\right)  =q\left(  M,\alpha,\beta\right)  .
\label{requaalq}%
\end{equation}

\subsection{\label{Section_error_T_discussion}The Eigen paradox}

Notice that by (\ref{decayq}), $r\left(  M,\alpha,\beta\right)  =q\left(
M,\alpha,\beta\right)  $ decays with $M$, so we can use the Maynard Smith
ansatz to estimate the error threshold, see Section
\ref{Section-Maynard-Smith}. From (\ref{error_thr}), by using \ that $q\left(
M,\alpha,\beta\right)  >\frac{\left(  p-1\right)  p^{-2M}}{C}e^{-\beta
p^{\alpha}}$, with $p$, $\alpha$, $\beta$, $s$ fixed, we have%
\[
\frac{\left(  p-1\right)  p^{-2M}}{C}e^{-\beta p^{\alpha}}<s,
\]
which implies that%
\begin{equation}
M>\frac{-\left(  \beta p^{\alpha}+\ln s\right)  }{2\ln p}+\frac{\ln\frac
{p-1}{C}}{2\ln p}. \label{key_ineqquality}%
\end{equation}
Under a `fierce competition'\ between the groups $I+p^{M}\mathbb{Z}_{p}$,
$\mathbb{Q}_{p}\smallsetminus\left[  I+p^{M}\mathbb{Z}_{p}\right]  $, i.e.
when rate $b$ approaches from the left to rate $a$ (i.e. $s\rightarrow0^{+}$),
$M$ must grow, which means that the survival of the sequences in the group
$I+p^{M}\mathbb{Z}_{p}$ demands that they get closer to master sequence $I$,
which means, that they must increase their lenghts. Then in our model \ the
`classical Eigen's paradox does not occur' because the length of the genomes
can grow during the evolution process. Notice that if $M$ does not satisfy
(\ref{key_ineqquality}), then the sequences in the set $I+p^{M}\mathbb{Z}_{p}$
will not survive in the long term.

For arbitrary $f$ and $Q$, \ we propose the existence of threshold function
for the length of the genomes $M_{c}(f,Q)$ such that $M>M_{c}(f,Q)$ is a
necessary and sufficient condition for the long term survival of the genomes.
Formula (\ref{key_ineqquality}) gives an estimate for the function
$M_{c}(f,Q)$ for the particular case in which $Q$ has the form
(\ref{EC_Q_form}). We propose that under the condition $M>M_{c}(f,Q)$%
,\ \textit{the Darwin-Eigen cycle}\ proposed by A. Poole, D. Jeffares and D.
Penny takes place, see \cite{Poole etal}: the Darwin-Eigen cycle is a positive
feedback mechanism. Larger genome size improves the fidelity replication
($q\left(  M,\alpha,\beta\right)  $ decays when $M$ grows cf. (\ref{decayq})),
and this increases the Eigen limit (see (\ref{key_ineqquality})) on the length
of the genome, and this allows the evolution of larger genome size. In turn,
this allows the evolution of new function, which could further improve the
replication fidelity, and so on. See \cite{Poole etal}, \cite{Sch}, and
\cite{James et al}, for a detailed biological discussion. On the other hand,
if $M\leq M_{c}(f,Q)$ then genomes do not survive in the long term. This is a
type of classical error threshold, `similar' to the one provided by the Eigen
evolution model with point mutation matrices.

\section{\label{Section_Cauchy_problem}The Cauchy Problem for the $p$-adic
replicator equation}

In this section we show the existence of a solution for the Cauchy problem
(\ref{eq2}), which in turn implies the existence of \textit{quasispecies} for
the $p$-adic replicator equation introduced here. This goal is achieve by
using the classical method of separation of variables and $p$-adic wavelets,
see \cite{Alberio et al} and \cite{KKZuniga}. We assume that the function $Q$
satisfies only the hypotheses given in Section \ref{Section_mutation_measure}.

\subsection{Some remarks on $p$-adic wavelets}

We take $K=\mathbb{C}$, $\mathbb{R}$. We denote by $C(\mathbb{Q}%
_{p},\mathbb{K})$ the $\mathbb{K}$-vector space of \ continuous $\mathbb{K}%
$-valued functions defined on $\mathbb{Q}_{p}$.

We fix a function $\mathfrak{a}:\mathbb{R}_{+}\rightarrow\mathbb{R}_{+}$ and
define the pseudodifferential operator%
\[%
\begin{array}
[c]{ccc}%
\mathcal{D} & \rightarrow & C(\mathbb{Q}_{p},\mathbb{C})\cap L^{2}\\
&  & \\
\varphi & \rightarrow & \boldsymbol{A}\varphi,
\end{array}
\]
where $\left(  \boldsymbol{A}\varphi\right)  \left(  x\right)  =\mathcal{F}%
_{\xi\rightarrow x}^{-1}\left\{  \mathfrak{a}\left(  \left\vert \xi\right\vert
_{p}\right)  \mathcal{F}_{x\rightarrow\xi}\varphi\right\}  $.

The set of functions $\left\{  \Psi_{rnj}\right\}  $ defined as%
\begin{equation}
\Psi_{rnj}\left(  x\right)  =p^{\frac{-r}{2}}\chi_{p}\left(  p^{r-1}jx\right)
\Omega\left(  \left\vert p^{r}x-n\right\vert _{p}\right)  , \label{eq4}%
\end{equation}
where $r\in\mathbb{Z}$, $j\in\left\{  1,\cdots,p-1\right\}  $, and $n$ runs
through a fixed set of representatives of $\mathbb{Q}_{p}/\mathbb{Z}_{p}$, is
an orthonormal basis of $L^{2}(\mathbb{Q}_{p},\mathbb{C})$ consisting of
eigenvectors of operator $\boldsymbol{A}$:%
\begin{equation}
\boldsymbol{A}\Psi_{rnj}=\mathfrak{a}(p^{1-r})\Psi_{rnj}\text{ for any
}r\text{, }n\text{, }j\text{.} \label{eq5}%
\end{equation}
This result is due to S. Kozyrev see e.g. \cite[Theorem 3.29]{KKZuniga},
\cite[Theorem 9.4.2]{Alberio et al}.\ Notice that%
\[
\widehat{\Psi}_{rnj}\left(  \xi\right)  =p^{\frac{r}{2}}\chi_{p}\left(
p^{-r}n\left(  \xi+p^{r-1}j\right)  \right)  \Omega\left(  \left\vert
p^{-r}\xi+p^{-1}j\right\vert _{p}\right)  ,
\]
and then%
\[
\mathfrak{a}\left(  \left\vert \xi\right\vert _{p}\right)  \widehat{\Psi
}_{rnj}\left(  \xi\right)  =\mathfrak{a}(p^{1-r})\widehat{\Psi}_{rnj}\left(
\xi\right)  .
\]

The Fourier transform $\widehat{Q}$ of $Q$ is a real-valued function, which is
radial in $\mathbb{Q}_{p}\smallsetminus\left\{  0\right\}  $. For this reason,
we use the notation $\widehat{Q}\left(  \left\vert \xi\right\vert _{p}\right)
$.

We set%
\[
\boldsymbol{W}\varphi=Q\ast\varphi\text{ for }\varphi\in\mathcal{D},
\]
as before. Then%
\begin{equation}
\boldsymbol{W}\Psi_{rnj}\left(  x\right)  =\widehat{Q}\left(  p^{1-r}\right)
\Psi_{rnj}\left(  x\right)  ,\label{nota_op_W}%
\end{equation}
where $\widehat{Q}\left(  p^{1-r}\right)  $ is a real number satisfying
$\left\vert \widehat{Q}\left(  p^{1-r}\right)  \right\vert \leq1$.

\subsection{The Cauchy problem for operator $\boldsymbol{W}$}

We now consider the following initial value problem:%
\begin{equation}
\left\{
\begin{array}
[c]{ll}%
Y:\mathbb{Q}_{p}\times\mathbb{R}_{+}\rightarrow\mathbb{R}\text{,} & Y\left(
\cdot,t\right)  \in L_{\mathbb{R}}^{2}\text{, }Y\left(  x,\cdot\right)  \in
C^{1}\left(  \mathbb{R}_{+},\mathbb{R}\right) \\
& \\
\frac{\partial Y\left(  x,t\right)  }{\partial t}=\boldsymbol{W}Y\left(
x,t\right)  , & x\in\mathbb{Q}_{p},t>0\\
& \\
Y\left(  x,0\right)  =Y_{0}\left(  x\right)  \in\mathcal{D}_{\mathbb{R}}. &
\end{array}
\right.  \label{eq6}%
\end{equation}
Notice that the conditions $Y_{0}\left(  x\right)  \geq0$ and $\int
_{\mathbb{Q}_{p}}Y_{0}\left(  x\right)  dx=1$ constitute natural physical
restrictions for the function $Y_{0}$, however, here we do not use them.

We solve (\ref{eq6}) by using the separation of variables method. We first
look for a complex-valued solution of (\ref{eq6}) of the form%
\begin{align}
\widetilde{Y}\left(  x,t\right)   &  =\sum\limits_{rjn}C_{rjn}\left(
t\right)  \Psi_{rnj}\left(  x\right) \label{eq6A}\\
&  =\sum\limits_{rjn}C_{rjn}\left(  t\right)  p^{\frac{-r}{2}}\chi_{p}\left(
p^{r-1}jx\right)  \Omega\left(  \left\vert p^{r}x-n\right\vert _{p}\right)
,\nonumber
\end{align}
where $C_{rjn}\left(  t\right)  $ are complex-valued functions, which admit
continuous temporal derivatives. We fix a countable disjoint covering of
$\mathbb{Q}_{p}$ by balls of the form:%
\[
B_{r_{i}}\left(  p^{-r_{i}}n_{i}\right)  =p^{-r_{i}}n_{i}+p^{-r_{i}}%
\mathbb{Z}_{p}\text{, }%
\]
where $r_{i}\in\mathbb{Z}$, and $n_{i}=a_{-k_{i}}p^{-k_{i}}+\cdots
+a_{-1}p^{-1}\in\mathbb{Q}_{p}/\mathbb{Z}_{p}$, and the digits $a_{j}$s runs
through the set $\left\{  0,\cdots,p-1\right\}  $, such that%
\begin{equation}
f\mid_{B_{r_{i}}\left(  p^{-r_{i}}n_{i}\right)  }=f(\left\vert p^{-r_{i}}%
n_{i}\right\vert _{p}). \label{Eq_condtion_fitness}%
\end{equation}
\ Consequently%
\begin{equation}
f\left(  \left\vert x\right\vert _{p}\right)  =\sum\limits_{i=0}^{l\left(
f\right)  }f(\left\vert p^{-r_{i}}n_{i}\right\vert _{p})\Omega\left(
\left\vert p^{r_{i}}x-n_{i}\right\vert _{p}\right)  \text{, }l(f)\in
\mathbb{N}\text{,} \label{Eq_fitness_function}%
\end{equation}
due to the fact that the fitness function $f$ is a locally constant function
with compact support.

We now describe the balls contained in $B_{r_{i}}\left(  p^{-r_{i}}%
n_{i}\right)  $. Any such ball has the form:%
\begin{gather}
a_{-k_{i}}p^{-r_{i}-k_{i}}+\cdots+a_{-1}p^{-r_{i}-1}+b_{1}p^{-r_{i}}%
+\cdots+b_{r-1}p^{r-1}+p^{r}\mathbb{Z}_{p}\nonumber\\
=p^{-r_{i}}n_{i}+p^{r}\left(  b_{1}p^{-r-r_{i}}+\cdots+b_{r-1}p^{-1}\right)
+p^{r}\mathbb{Z}_{p}=:p^{-r_{i}}n_{i}+p^{r}n_{r}+p^{r}\mathbb{Z}%
_{p}\nonumber\\
=p^{r}\left(  p^{-r-r_{i}}n_{i}+n_{r}\right)  +p^{r}\mathbb{Z}_{p},
\label{Eq_balls_0}%
\end{gather}
for some integer $r\geq-r_{i}$ and some $n_{r}\in\mathbb{Q}_{p}/\mathbb{Z}%
_{p}$. \ The amount of such balls is exactly $p^{r-1}$. Now, for a fixed
$\Omega\left(  \left\vert p^{r_{i}}x-n_{i}\right\vert _{p}\right)  $ and a
variable $\Omega\left(  \left\vert p^{r}x-n\right\vert _{p}\right)  $, we
have
\begin{equation}
\Omega\left(  \left\vert p^{r}x-n\right\vert _{p}\right)  \Omega\left(
\left\vert p^{r_{i}}x-n_{i}\right\vert _{p}\right)  =\left\{
\begin{array}
[c]{ll}%
\Omega\left(  \left\vert p^{r}x-n\right\vert _{p}\right)  & \text{if }%
B_{r}\left(  p^{-r}n\right)  \subset B_{r_{i}}\left(  p^{-r_{i}}n_{i}\right)
\\
& \\
0 & \text{if }B_{r}\left(  p^{-r}n\right)  \nsubseteqq B_{r_{i}}\left(
p^{-r_{i}}n_{i}\right)  .
\end{array}
\right.  \label{Eq_balls}%
\end{equation}
By using (\ref{Eq_balls_0})-(\ref{Eq_balls}), we have%
\begin{align*}
f\left(  \left\vert x\right\vert _{p}\right)  \widetilde{Y}\left(  x,t\right)
&  =\sum\limits_{i=0}^{l(f)}f(\left\vert p^{-r_{i}}n_{i}\right\vert _{p}%
)\sum\limits_{j=1}^{p-1}\sum\limits_{r\geq-r_{i}}\text{ }\sum
\limits_{p^{-r-r_{i}}n_{i}+n_{r}}C_{\left(  -r\right)  j\left(  p^{-r-r_{i}%
}n_{i}+n_{r}\right)  }\left(  t\right)  \times\\
&  p^{\frac{r}{2}}\chi_{p}\left(  p^{-r-1}jx\right)  \Omega\left(  \left\vert
p^{-r}x-\left(  p^{-r-r_{i}}n_{i}+n_{r}\right)  \right\vert _{p}\right)  ,
\end{align*}
and by using (\ref{nota_op_W}),%
\begin{gather}
\boldsymbol{W}\left(  f\left(  \left\vert x\right\vert _{p}\right)
\widetilde{Y}\left(  x,t\right)  \right)  =\label{eq8}\\
\sum\limits_{i=0}^{l(f)}f(\left\vert p^{-r_{i}}n_{i}\right\vert _{p}%
)\sum\limits_{j=1}^{p-1}\sum\limits_{r\geq-r_{i}}\text{ }\sum
\limits_{p^{-r-r_{i}}n_{i}+n_{r}}C_{\left(  -r\right)  j\left(  p^{-r-r_{i}%
}n_{i}+n_{r}\right)  }\left(  t\right)  \times\nonumber\\
\widehat{Q}(p^{1+r})p^{\frac{r}{2}}\chi_{p}\left(  p^{-r-1}jx\right)
\Omega\left(  \left\vert p^{-r}x-\left(  p^{-r-r_{i}}n_{i}+n_{r}\right)
\right\vert _{p}\right)  .\nonumber
\end{gather}
By replacing (\ref{eq6A}) and (\ref{eq8}) in (\ref{eq6}), we obtain
\begin{gather}
\widetilde{Y}\left(  x,t\right)  =\sum\limits_{i=0}^{l(f)}f(\left\vert
p^{-r_{i}}n_{i}\right\vert _{p})\sum\limits_{j=1}^{p-1}\sum\limits_{r\geq
-r_{i}}\text{ }\sum\limits_{p^{-r-r_{i}}n_{i}+n_{r}}C_{\left(  -r\right)
j\left(  p^{-r-r_{i}}n_{i}+n_{r}\right)  }\left(  0\right)  \times
\label{Eq_Y_tilde}\\
\exp\left(  f(\left\vert p^{-r_{i}}n_{i}\right\vert _{p})\widehat{Q}%
(p^{1+r})t\right)  p^{\frac{r}{2}}\chi_{p}\left(  p^{-r-1}jx\right)
\Omega\left(  \left\vert p^{-r}x-\left(  p^{-r-r_{i}}n_{i}+n_{r}\right)
\right\vert _{p}\right)  ,\nonumber
\end{gather}
where
\begin{align*}
Y_{0}\left(  x\right)   &  =\sum\limits_{i=0}^{l(f)}f(\left\vert p^{-r_{i}%
}n_{i}\right\vert _{p})\sum\limits_{j=1}^{p-1}\sum\limits_{r\geq-r_{i}}\text{
\ }\sum\limits_{p^{-r-r_{i}}n_{i}+n_{r}}C_{\left(  -r\right)  j\left(
p^{-r-r_{i}}n_{i}+n_{r}\right)  }\left(  0\right)  \times\\
&  p^{\frac{r}{2}}\chi_{p}\left(  p^{-r-1}jx\right)  \Omega\left(  \left\vert
p^{-r}x-\left(  p^{-r-r_{i}}n_{i}+n_{r}\right)  \right\vert _{p}\right)  .
\end{align*}
Now, we set
\[
\alpha_{rr_{i}n_{i}}:=f(\left\vert p^{-r_{i}}n_{i}\right\vert _{p})\widehat
{Q}(p^{1+r}),
\]
then%
\begin{gather*}
\exp\left[  f(\left\vert p^{r_{i}}n_{i}\right\vert _{p})\widehat{Q}%
(p^{1+r})t+2\pi i\left\{  p^{-r-1}jx\right\}  _{p}\right]  =\\
e^{t\alpha_{rr_{i}n_{i}}}\left[  \cos\left(  2\pi\left\{  p^{-r-1}jx\right\}
_{p}\right)  +\sqrt{-1}\sin\left(  2\pi\left\{  p^{-r-1}jx\right\}
_{p}\right)  \right]  ,
\end{gather*}
and by setting%
\begin{align*}
A_{rjr_{i}n_{i}n_{r}}  &  :=\operatorname{Re}C_{\left(  -r\right)  j\left(
p^{-r-r_{i}}n_{i}+n_{r}\right)  }\left(  0\right)  \text{, }\\
B_{rjr_{i}n_{i}n_{r}}  &  :=\operatorname{Im}C_{\left(  -r\right)  j\left(
p^{-r-r_{i}}n_{i}+n_{r}\right)  }\left(  0\right)  \text{,}%
\end{align*}
the solution $Y(x,t)$ is the real part of $\widetilde{Y}\left(  x,t\right)  $:%
\begin{gather}
Y(x,t)=\sum\limits_{i=0}^{l(f)}\sum\limits_{j=1}^{p-1}\sum\limits_{r\geq
-r_{i}}\text{ }\sum\limits_{p^{-r-r_{i}}n_{i}+n_{r}}p^{\frac{r}{2}}%
e^{t\alpha_{rr_{i}n_{i}}}\Omega\left(  \left\vert p^{-r}x-\left(  p^{-r-r_{i}%
}n_{i}+n_{r}\right)  \right\vert _{p}\right)  \times\label{eq9}\\
\left[  A_{rjr_{i}n_{i}n_{r}}\cos\left(  2\pi\left\{  p^{-r-1}jx\right\}
_{p}\right)  -B_{rjr_{i}n_{i}n_{r}}\sin\left(  2\pi\left\{  p^{-r-1}%
jx\right\}  _{p}\right)  \right]  .\nonumber
\end{gather}
In the next section we show that $Y(x,t)$ is really a finite sum, if $Y_{0}%
\in\mathcal{D}_{\mathbb{R}}$.

\subsubsection{\label{note1}Some remarks about the Fourier coefficients}

If $Y_{0}\in\mathcal{D}_{\mathbb{R}}$, then
\[
C_{\left(  -r\right)  j\left(  p^{-r-r_{i}}n_{i}+n_{r}\right)  }\left(
0\right)  \neq0\Leftrightarrow-r_{i}\leq r\leq l-1,
\]
where $l$ is the index of local constancy of $Y_{0}$. This implies that almost
all the $C_{\left(  -r\right)  j\left(  p^{-r-r_{i}}n_{i}+n_{r}\right)
}\left(  0\right)  $s are zero.

Indeed,%
\[
f(\left\vert p^{-r_{i}}n_{i}\right\vert _{p})C_{\left(  -r\right)  j\left(
p^{-r-r_{i}}n_{i}+n_{r}\right)  }\left(  0\right)  =p^{\frac{r}{2}}%
\int\limits_{p^{-r_{i}}n_{i}+p^{r}n_{r}+p^{r}\mathbb{Z}_{p}}Y_{0}\left(
x\right)  \chi_{p}\left(  p^{-r-1}jx\right)  dx\text{.}%
\]
We now use the subdivision%
\[
p^{-r_{i}}n_{i}+p^{r}n_{r}+p^{r}\mathbb{Z}_{p}=\bigsqcup\limits_{n_{l}%
}p^{-r_{i}}n_{i}+p^{r}n_{r}+p^{l}n_{l}+p^{l}\mathbb{Z}_{p}.
\]
Then $f(\left\vert p^{-r_{i}}n_{i}\right\vert _{p})C_{\left(  -r\right)
j\left(  p^{-r-r_{i}}n_{i}+n_{r}\right)  }\left(  0\right)  $\ becomes a
finite sum of terms of the form%
\begin{gather*}
Y_{0}\left(  p^{-r_{i}}n_{i}+p^{r}n_{r}+p^{l}n_{l}\right)  \int
\limits_{p^{-r_{i}}n_{i}+p^{r}n_{r}+p^{l}n_{l}+p^{l}\mathbb{Z}_{p}}\chi
_{p}\left(  p^{-r-1}jx\right)  dx\\
=p^{-l}Y_{0}\left(  p^{-r_{i}}n_{i}+p^{r}n_{r}+p^{l}n_{l}\right)  \chi
_{p}\left(  p^{-r-1}j\left\{  p^{-r_{i}}n_{i}+p^{r}n_{r}+p^{l}n_{l}\right\}
\right)  \times\\
\int\limits_{\mathbb{Z}_{p}}\chi_{p}\left(  p^{l-r-1}jy\right)  dy\\
=p^{-l}Y_{0}\left(  p^{-r_{i}}n_{i}+p^{r}n_{r}+p^{l}n_{l}\right)  \chi
_{p}\left(  p^{-r-1}j\left\{  p^{-r_{i}}n_{i}+p^{r}n_{r}+p^{l}n_{l}\right\}
\right)  \times\\
\left\{
\begin{array}
[c]{ccc}%
1 & \text{if} & r\leq l-1\\
&  & \\
0 & \text{if} & r>l-1.
\end{array}
\right.
\end{gather*}
For details about the calculation of the integral involving the additive
character, the reader may consult for instance \cite[p. 42-43]{V-V-Z}.

\subsection{The Cauchy Problem for the $p$-adic replicator equation}

We now consider the following Cauchy problem:%
\begin{equation}
\left\{
\begin{array}
[c]{ll}%
X:\mathbb{Q}_{p}\times\mathbb{R}_{+}\rightarrow\mathbb{R}\text{,} & X\left(
\cdot,t\right)  \in L_{\mathbb{R}}\cap L_{\mathbb{R}}^{2}\text{, }X\left(
x,\cdot\right)  \in C^{1}\left(  \mathbb{R}_{+},\mathbb{R}\right) \\
& \\
\frac{\partial X\left(  x,t\right)  }{\partial t}=\boldsymbol{W}X\left(
x,t\right)  -\Phi\left(  t\right)  X\left(  x,t\right)  , & x\in\mathbb{Q}%
_{p},t>0\\
& \\
X\left(  x,0\right)  =Y_{0}\in\mathcal{D}_{\mathbb{R}}. &
\end{array}
\right.  \label{Cauchy_p_replicator_equation}%
\end{equation}
By using (\ref{hypotesis_1}), (\ref{eq1A}) and (\ref{eq9}), we have
\begin{gather*}
\exp\left(  \int_{0}^{t}\Phi\left(  \tau\right)  d\tau\right)  =\int
\limits_{\mathbb{Q}_{p}}Y\left(  x,t\right)  dx\\
=\sum\limits_{i=0}^{l(f)}\sum\limits_{j=1}^{p-1}\sum\limits_{r\geq-r_{i}%
}\text{ }\sum\limits_{p^{-r-r_{i}}n_{i}+n_{r}}p^{\frac{r}{2}}e^{t\alpha
_{rr_{i}n_{i}}}D_{rjr_{i}n_{i}n_{r}}=:\overline{Y}\left(  t\right)  ,
\end{gather*}
where the $D_{rjr_{i}n_{i}n_{r}}$s are real constants, which are obtained by
integrating (\ref{eq9}) termwise with respect to $x$.

Therefore the solution of initial value problem
(\ref{Cauchy_p_replicator_equation}) is given by
\begin{gather}
X(x,t)=\frac{1}{\overline{Y}\left(  t\right)  }\sum\limits_{i=0}^{l(f)}%
\sum\limits_{j=1}^{p-1}\sum\limits_{r\geq-r_{i}}\text{ }\sum
\limits_{p^{-r-r_{i}}n_{i}+n_{r}}p^{\frac{r}{2}}e^{t\alpha_{rr_{i}n_{i}}%
}\Omega\left(  \left\vert p^{-r}x-\left(  p^{-r-r_{i}}n_{i}+n_{r}\right)
\right\vert _{p}\right) \label{eq10}\\
\times\left[  A_{rjr_{i}n_{i}n_{r}}\cos\left(  2\pi\left\{  p^{-r-1}%
jx\right\}  _{p}\right)  -B_{rjr_{i}n_{i}n_{r}}\sin\left(  2\pi\left\{
p^{-r-1}jx\right\}  _{p}\right)  \right]  .\nonumber
\end{gather}

Notice that $\int_{\mathbb{Q}_{p}}X(x,t)dx=1$ for $t>0$, and thus the
hypothesis (\ref{hypotesis_1}) holds, and consequently $X\left(
\cdot,t\right)  \in L_{\mathbb{R}}$ for $t>0$.

\subsection{The $p$-adic quasispecies}

In this section we consider the steady state concentration:%
\begin{equation}
\overline{X}(x):=\lim_{t\rightarrow+\infty}X(x,t), \label{eq_stationary_con}%
\end{equation}
which corresponds, in the classical terminology, to \textit{the }%
$p$\textit{-adic quasispecies}. We define%
\[
\lambda_{\max}=\max_{\substack{0\leq i\leq l\left(  f\right)  \\r\geq-r_{i}%
}}\left\{  f(\left\vert p^{-r_{i}}n_{i}\right\vert _{p})\widehat{Q}%
(p^{1+r})\right\}  .
\]
We recall that the condition $r\geq-r_{i}$ involves only a finite number of
$r$s, since \ all the sums in (\ref{eq10}) run through finite sets, this is a
consequence of the fact that the Fourier expansion of $\widetilde{Y}\left(
x,t\right)  $\ is finite. We now define%
\[
T=\left\{  rr_{i}n_{i};f(\left\vert p^{-r_{i}}n_{i}\right\vert _{p}%
)\widehat{Q}(p^{1+r})=\lambda_{\max}\right\}  .
\]
Then, we have
\begin{gather}
\overline{X}(x)=\frac{1}{C}\sum\limits_{rr_{i}n_{i}\in T}\text{ }%
\sum\limits_{j=1}^{p-1}p^{\frac{r}{2}}\Omega\left(  \left\vert p^{-r}x-\left(
p^{-r-r_{i}}n_{i}+n_{r}\right)  \right\vert _{p}\right)  \times
\label{Eq_X_bar}\\
\left[  A_{rjr_{i}n_{i}n_{r}}\cos\left(  2\pi\left\{  p^{-r-1}jx\right\}
_{p}\right)  -B_{rjr_{i}n_{i}n_{r}}\sin\left(  2\pi\left\{  p^{-r-1}%
jx\right\}  _{p}\right)  \right]  ,\nonumber
\end{gather}
where%
\[
C=\sum\limits_{rr_{i}n_{i}\in T}\sum\limits_{j=1}^{p-1}p^{\frac{r}{2}%
}D_{rjr_{i}n_{i}n_{r}}.
\]

\begin{notation}
If $U$ is an open and compact subset of $\mathbb{Q}_{p}$, \ for instance a
finite union of balls, we denote by $\mathcal{D}_{\mathbb{R}}(U)$, the
$\mathbb{R}$-vector space of test functions with supports in $U$.
\end{notation}

Consider the following eigenvalue problem:%
\begin{equation}
\boldsymbol{W}\varphi=\lambda\varphi\text{, for }\lambda\in\mathbb{R}\text{,
}\varphi\in\mathcal{D}_{\mathbb{R}}(\coprod\limits_{i=0}^{l(f)}B_{r_{i}%
}\left(  p^{-r_{i}}n_{i}\right)  ), \label{Eigenvalue_problem}%
\end{equation}
where $%
{\textstyle\coprod\nolimits_{i=0}^{l(f)}}
B_{r_{i}}\left(  p^{-r_{i}}n_{i}\right)  $ is the support of $f$, see
(\ref{Eq_fitness_function}). Then $\lambda_{\max}$\ is the largest eigenvalue
associated with (\ref{Eigenvalue_problem}).

On the other hand, from (\ref{Eq_X_bar}) we obtain that $\overline{X}(x)$ is a
continuous function with compact support, consequently, this function is
integrable. By using the dominated convergence theorem and condition
(\ref{hypotesis_1}), we have%
\[
1=\lim_{t\rightarrow+\infty}\int\nolimits_{\mathbb{Q}_{p}}X\left(  x,t\right)
dx=\int\nolimits_{\mathbb{Q}_{p}}\lim_{t\rightarrow+\infty}X\left(
x,t\right)  dx=\int\nolimits_{\mathbb{Q}_{p}}\overline{X}\left(  x\right)
dx.
\]
We now use the fact that $X\left(  x,t\right)  \geq0$ (due to the physical
meaning of this function), and thus $\overline{X}\left(  x\right)  \geq0$, in
this way we reach the conclusion that%
\[
\overline{X}\left(  x\right)  \text{ is a probability density supported in
supp }f\text{.}%
\]
This behavior is completely different from the one presented in the Eigen
model. \ In the classical case, under the hypothesis that the mutation matrix
has one largest eigenvalue, the steady state concentration is a constant
vector having exactly one non-zero entry.

\begin{acknowledgement}
The author wishes to thank the referees for their careful reading of the
original manuscript and for their helpful suggestions.
\end{acknowledgement}

\end{document}